\documentclass[12pt,preprint]{aastex}
 \newcommand{\beq}{\begin{equation}}
 \newcommand{\eeq}{\end{equation}}
 \newcommand{\beqn}{\begin{eqnarray}}
 \newcommand{\eeqn}{\end{eqnarray}}

\begin{document}
\title{
A comprehensive analysis of NGC 2158 in Gaia era: photometric parameters, apex and orbit}
\author{Devesh P. Sariya$^1$,
Ing-Guey Jiang$^1$,
M. D. Sizova$^2$, 
E. S. Postnikova$^2$,
D. Bisht$^3$,
N. V. Chupina$^2$, 
S.V. Vereshchagin$^2$,
R. K. S. Yadav$^4$,
G. Rangwal$^5$, and
A. V. Tutukov$^2$
}
\affil{
{$^1$Department of Physics and Institute of Astronomy,}\\
{National Tsing-Hua University, Hsin-Chu, Taiwan}\\
{$^2$Institute of Astronomy Russian Academy of Sciences (INASAN),}\\ 
{48 Pyatnitskaya st., Moscow, Russia}\\
{$^3$Key Laboratory for Researches in Galaxies and Cosmology,}\\ 
{University of Science and Technology of China, Chinese
Academy of Sciences, Hefei, Anhui, 230026, China}\\
{$^4$Aryabhatta Research Institute of Observational Sciences,}\\ 
{Manora Peak, Nainital 263002, India}\\
{$^5$Center of Advanced Study, Department of Physics, D. S. B. Campus,}\\ 
{Kumaun University Nainital 263002, India}
}

\begin{abstract}

We present an investigation of NGC 2158 
using Gaia-DR2 data. 
We identified 800 most likely cluster members 
with membership probability higher than $90\%$.
The mean proper motions of this object are determined as 
($\mu_{x}=-0.203\pm0.003$, $\mu_{y}=-1.99\pm0.004$) mas yr$^{-1}$. 
The limiting radius, log(age), and distance of the cluster are obtained as
23.5 arcmin, 9.38$\pm$0.04 Gyr 
and $4.69\pm0.22$ kpc, respectively.
The overall mass function slope ($ 0.93\pm0.14$)
is flatter than the Salpeter value (1.35) within the mass
range 1.17$-$1.44 $M_\odot$.
This cluster also shows the mass-segregation effect 
and our study demonstrates that
NGC 2158 is a dynamically relaxed open cluster. 
Using the AD-diagram, the apex coordinates of the cluster are obtained 
in different ways and examined using ($\mu_{U}$,$\mu_{T}$) diagram.
The best value of apex coordinates is determined as  
$A$=87.24$^\circ\pm$1.60$^\circ$, $D$=$-$36.61$^\circ\pm$5.30$^\circ$.
We also determined the orbit of the cluster and found that
NGC 2158 moves almost in the Solar antapex direction.
The resulting spatial velocity of NGC 2158 is 51 km s$^{-1}$. 
A significant oscillations along the $Z$-coordinate up to 529~pc is detected.
Various scenarios regarding the origin of this cluster are also discussed.

\end{abstract}

\keywords{Star:-Color-Magnitude diagrams - open cluster and associations: individual: NGC 2158-astrometry-Kinematics-apex-orbit}

\section{Introduction}
\label{INTRO}

Open clusters (OCs) are stellar systems with diffuse morphology 
found in the Galactic disk.
Since the stars of an OC are born out of the same molecular cloud,
the stars share many properties including age, distance, 
metallicity and motion. Because of their positions 
in the Galactic disk and varying ages, these objects are 
outstanding tracers of the Galactic disk
(Yadav et al. 2011, 2013; Soubiran et al. 2018; 
Cantat-Gaudin et al. 2018; Bossini et al. 2019). 

Proper motions (PMs) 
of the stars in a cluster would be 
roughly the same (with some dispersion). This
property is used to separate field stars lying in the region of the 
clusters and produce a sample with most likely cluster members 
(Anderson et al. 2006;  Yadav et al. 2008, 2013; Bellini et al. 2009; 
Zloczewski et al. 2011; 
Sariya et al. 2012, 2017, 2018; Bisht et al. 2020).
Owing to its high precision, 
the data from the Gaia space based satellite 
has been extensively used
for the study of Galaxy and its star clusters
(Cantat-Gaudin et al. 2018, 2020; Elsanhoury et al. 2018;
Gao 2018; Soubiran et al. 2018; Bisht et al. 2019, 2020; Bossini et al. 2019; 
Kunkel 2020; Chrobakova 2020; Ferreira et al. 2020; Rangwal et al. 2019; Postnikova et al. 2020).

NGC~2158 (OCL~468, Lund~206, Melotte~40) is a low metallicity open cluster
(Mochejska et al. 2004), lying towards the Galactic anticenter (Christian et al. 1985)
in a direction half degree from M~35 (NGC~2168) in the Gemini constellation.
The cluster is positioned at a relatively high location
above the galactic plane near Z=110~pc (Carraro et al. 2002).
It was classified as a II3r cluster by Trumpler (1930).
Due to its richness, it was initially confused as a globular cluster,
but ultimately due to its diffuse core, it was classified as an open cluster
by Shapley (1930).
Several photometric studies of the cluster have done over the decades
(Christian et al. 1985; Arp \& Cuffey 1962; Janes 1979; Hardy 1981; Carraro et al. 2002).
Carraro et al. (2002) provided a distance value equal to 3600$\pm$400 pc and 
an age value of $\sim$2 Gyr. They also suggested that NGC 2158 
is a genuine member of the old thin disk population. 
This fact makes it an interesting object to get the implications about both 
Galactic evolution and stellar evolution theories. 
In addition to this, NGC 2158 is similar to the populous intermediate-age 
OCs in the Large Magellanic Clouds (Christian et al. 1985). 
Bedin et al. (2010) used HST data to reach the white dwarf 
cooling sequence of the cluster
and gave an age between $\sim$1.8 and $\sim$2.0 Gyr of the cluster.
The variable stars in the cluster were investigated by Nardiello et al. (2015) using the Kepler data.
The catalog by Kharchenko et al. (2013) listed the distance value of 
NGC 2158 as 4770 pc. 
The central coordinates of the cluster are 
($\alpha, \delta$, J2000) = ($06^h07^m26.88^s, +24^\circ 05' 56.4''$), 
and ($l, b$) = ($186^\circ.635, +1^\circ.788$) according to the recent catalog by Cantat-Gaudin et al. (2018).
They gave a distance value of 4535.1 pc for the cluster.
The mean PM value of NGC 2158 estimated by Cantat-Gaudin et al. (2018) is 
($\mu_{\alpha}cos\delta, \mu_{\delta} = -0.177,-2.002$ mas yr$^{-1}$). 
Recently, the blue stragglers of NGC 2158 were studied 
by Vaidya et al. (2020).

The kinematical data from Gaia can be used to determine the apex (vertex)
point of OCs. This is the point on the celestial sphere where
the space velocity vectors seem to converge
(Vereshchagin et al. 2014; Elsanhoury et al. 2016, 2018;
Postnikova et al. 2020; Bisht et al. 2020).

%NGC 2158 has an intermediate form between a globular and an open cluster. 
The location of NGC 2158 being in the Galactic anticenter direction 
on a large distance to the Sun has suggested this cluster could be accreted.
In this paper, we also investigate if this is a possible scenario 
behind the origin of the cluster.

Among the upcoming sections of this paper, Section~\ref{OBS}
describes the Gaia DR2 data we used. 
The separation of field stars is shown in Section~\ref{PM}
followed by the determination of membership probability in Section~\ref{MP}.
Various photometric parameters of NGC 2158 are derived in Section~\ref{FUNDA}.
The determination of the cluster apex is discussed in Section~\ref{adapex}.
The orbit of NGC 2158 in the Galaxy is obtained in Section~\ref{orbit}. 
Finally, the conclusions of this work are presented in Section~\ref{CON}.

\section{Data}
\label{OBS}

\begin{figure}
%\centering
\includegraphics[width=\textwidth]{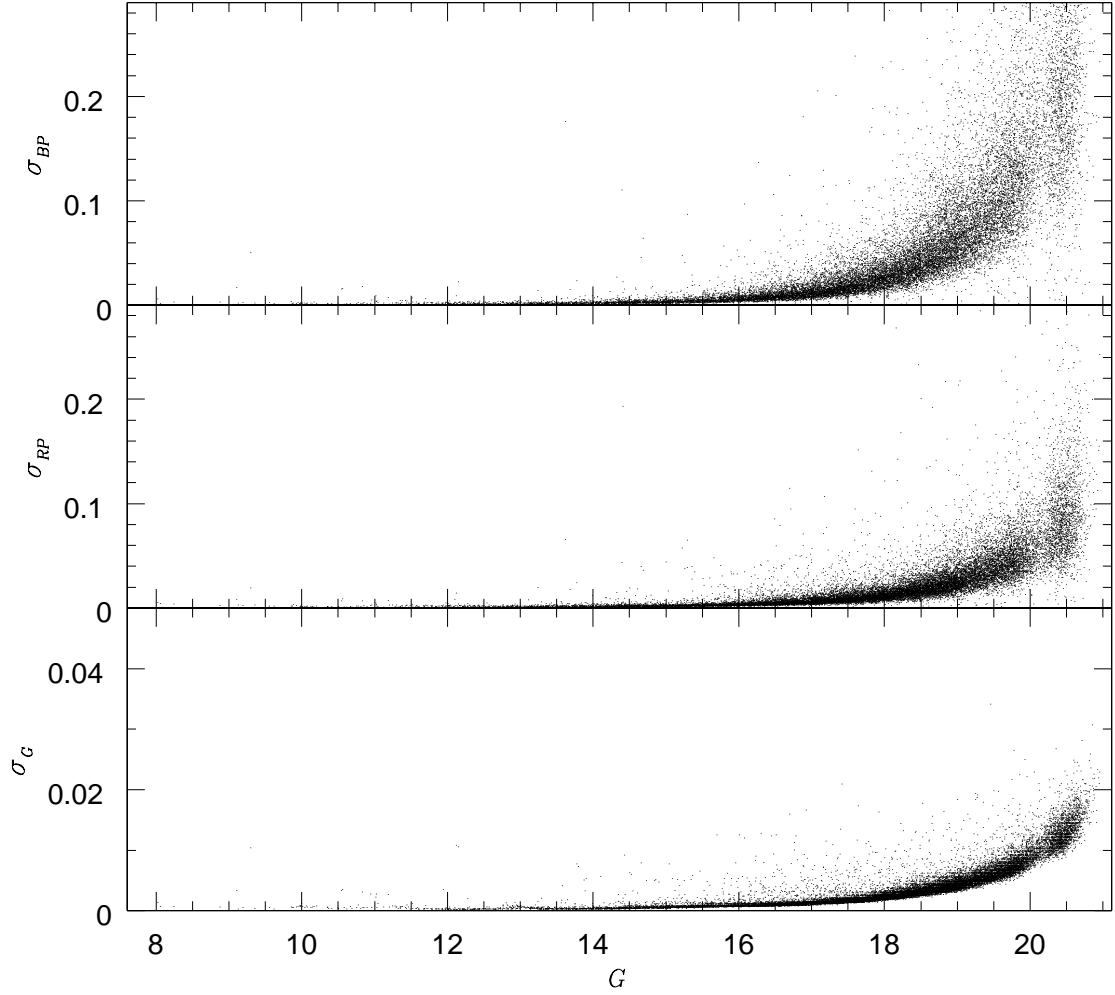}
\caption{
Errors of three magnitude bands with $G$ magnitude.
The errors in Gaia $G$ band are much better than the other two bands.
}
\label{figerrmag}
\end{figure}

\begin{figure}
%\centering
\includegraphics[width=\textwidth]{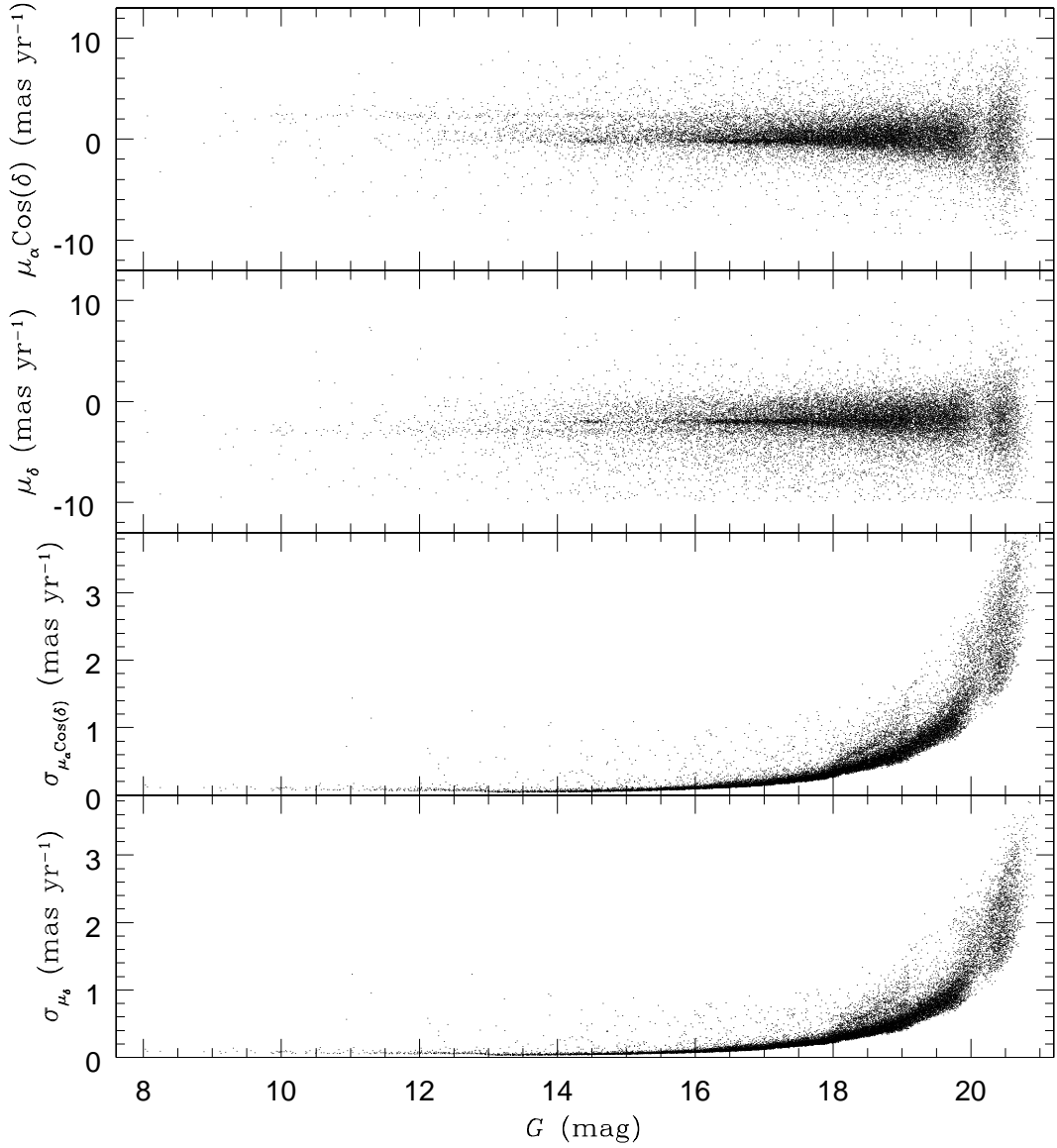}
\caption{
The distribution of Gaia DR2 PMs and their errors for NGC 2158 
versus $G$ magnitude.
}
\label{figerrpm}
\end{figure}

We used the Gaia data from their recent data release,
i.e. second Gaia Data Release 
(Gaia-DR2 or GDR2, Gaia Collaboration et al. 2018a,b).
GDR2 data is fully capable of an in-depth study of the cluster's
kinematics and dynamics, owing to a five parameter astrometric solution in the data.
These include celestial positions $(\alpha, \delta)$, parallaxes and proper motion
components ($\mu_{\alpha} cos\delta , \mu_{\delta}$).
The Gaia catalog also provides photometric information in three photometric bands
$G, G_{BP}$ and $G_{RP}$.
The main photometric band is the $G$ band that covers a wavelength range between 
330 to 1050 nm with limiting magnitude of $G=21$ mag.
For stars brighter than G = 17~mag, 
parallax error is $<0.1$~mas 
and PM error is $<0.2$~mas~yr$^{-1}$ (Lindegren et al. 2018).
For NGC 2158, the Gaia magnitude bands data are plotted with respect to $G$ magnitude
in Fig.~\ref{figerrmag}. 
The mean error in $G$ band is $\sim$0.0014 mag for stars 
brighter than 18~mag.
Similarly, for $18<G<20$ mag, 
the nominal uncertainties
reach $\sim$0.04~mag in $G_{RP}$ and $\sim$0.1~mag in $G_{BP}$.
We have plotted the PMs and their respective errors against
$G$ magnitude shown in Fig.~\ref{figerrpm} within a radius
of 25 arcmin from the center of NGC 2158. In this figure we did not apply a
membership criterion for any star. The median error in PM components is
$\sim$0.4 mas~yr$^{-1}$ for stars brighter than $G$=20~mag.

\section{Cluster CMD decontamination}
\label{PM}

\begin{figure}
%\centering
\includegraphics[width=\textwidth]{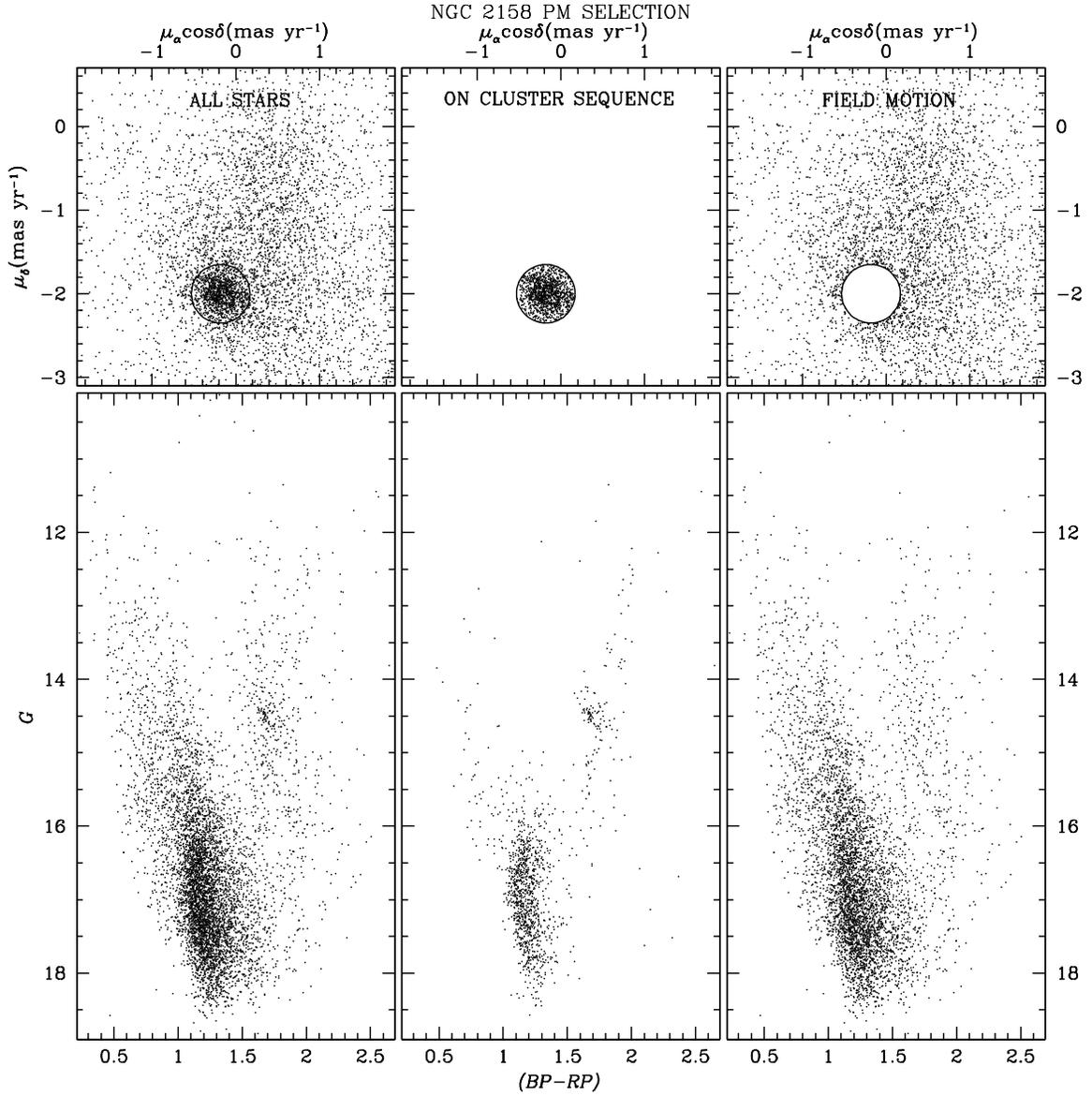}
\caption{
Shown here are the VPDs (top panels) and CMDs (bottom panels) 
using the Gaia-DR2 data for NGC 2158.
The circle of radius value $\sim$0.35 mas~yr$^{-1}$ in the VPDs represent
the provisionally assumed cluster stars.
The left side VPD and CMD show the entire sample of stars.
The central panels are for the cluster stars and 
the field star population is shown in the right panels. 
In this diagram, only the stars with PM errors less 1.0 mas~yr$^{-1}$
and parallax error less than 1.0 mas are plotted. 
}
\label{figvpd}
\end{figure}

For a cluster, the proper motion components 
($\mu_{\alpha} cos{\delta}$, $\mu{\delta}$) 
can be plotted as a vector point diagram (VPD)
(Bellini et al. 2009; Sariya et al. 2017, 2018)
The VPD for NGC 2158 is shown in Fig.~\ref{figvpd}. 
As can be seen in the top panels of the figure, 
a strong concentration shows the similar motion of stars
belonging to the NGC 2158,
while the scattered distribution shows field stars.
%The radius of the circle shown in VPD is decided by its effect 
%on the cluster sequences of CMD.
The left panels show all the stars, the middle panels show 
only the cluster stars and 
the right panels show the separated field stars. For plotting 
the Fig.~\ref{figvpd}, 
we have included stars with 
PM error $<$ 1.0 mas yr$^{-1}$ and parallax error $<$ 1.0 mas. 

\section{Membership probabilities}
\label{MP}

One of the primary applications of PMs in the direction of a star
cluster is to determine membership probabilities 
(Cudworth 1986; Jones 1997).
The membership probability can be used to select a sample of
most probable cluster members (Sariya et al. 2015, 2018; Bisht et al. 2020).
Here, we use the method given by Balaguer-N\'{u}\~{n}ez et al. (1998)
to calculate the membership probabilities in the region of NGC 2158.
This method has been well explained in detail by Bisht et al. (2020).

The stars chosen for the membership calculation have their 
PM errors better than 1.0 mas~yr$^{-1}$ and 
the parallax errors are $<$ 1.0 mas . 
The VPD circle of radius 0.35 mas~yr$^{-1}$
is selected for the preliminary cluster membership. 
The PM centers for the cluster motion is 
($\mu_{xc}$, $\mu_{yc}$) = ($-$0.177, $-$1.998) mas~yr$^{-1}$ .
For the field motion, the values of ($\mu_{\alpha}cos\delta, \mu_{\delta}$)
are noted as ($\mu_{xf}$, $\mu_{yf}$) by Balaguer-N\'{u}\~{n}ez et al. (1998).
Here, we have ($\mu_{xf}$, $\mu_{yf}$)= (0.314,$-$1.646) mas~yr$^{-1}$.
The PM dispersion for cluster stars is $\sigma_c$ = 0.1 mas~yr$^{-1}$
and the dispersion in the field motion is 
($\sigma_{xf}$, $\sigma_{yf}$) = (0.6,0.9) mas~yr$^{-1}$.

\begin{figure}
%\centering
\includegraphics[width=\textwidth]{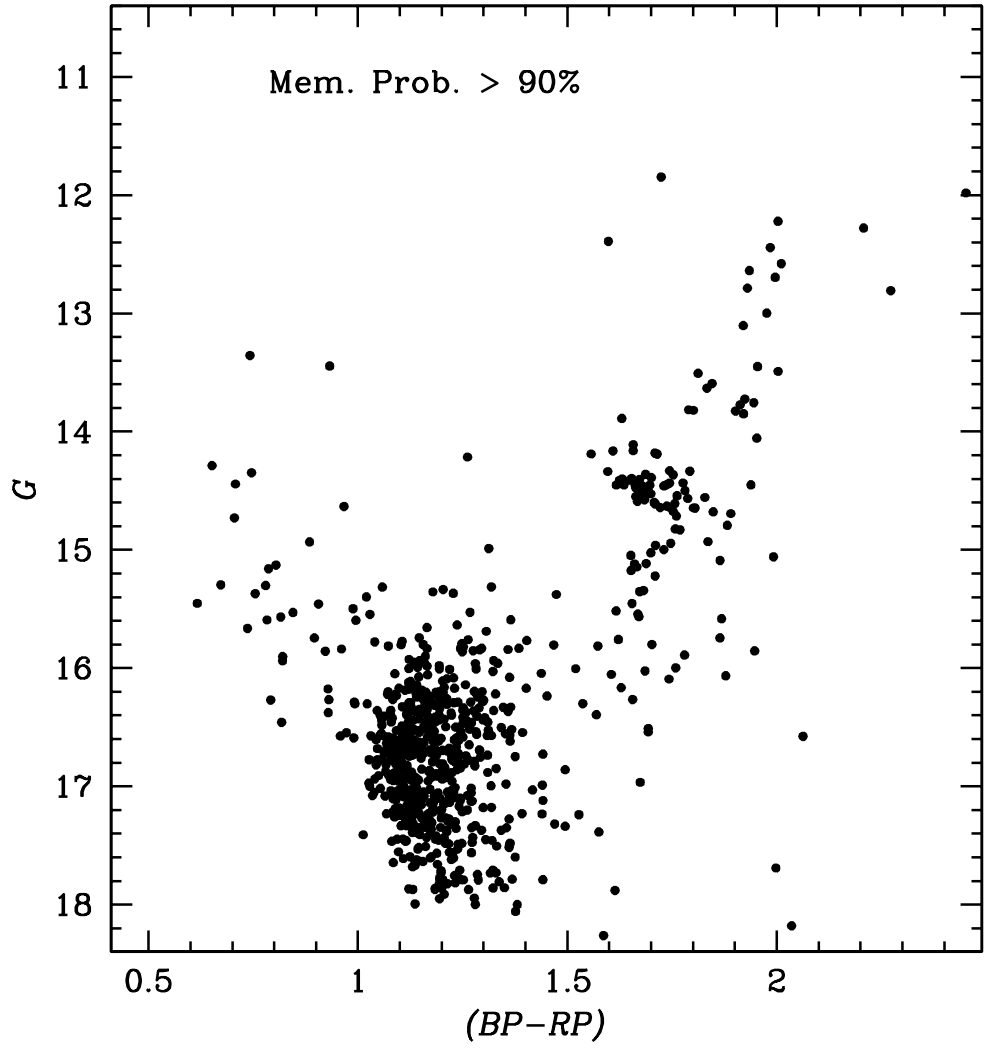}
\caption{
CMD of the most probable cluster members of NGC 2158. 
These stars have their membership probability value greater than 90\%.
}
\label{figmp90}
\end{figure}

\begin{figure}
%\centering
\includegraphics{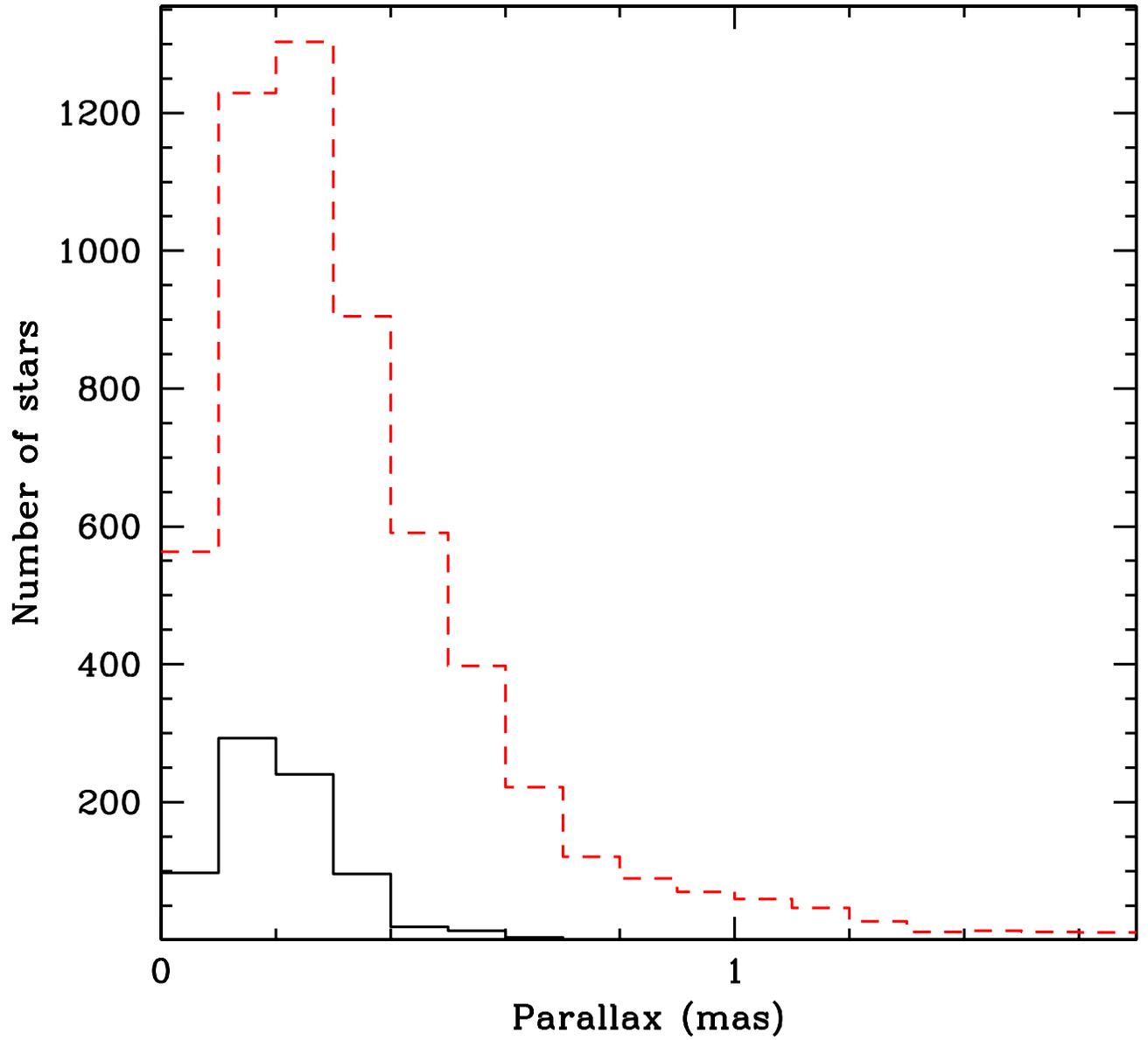}
\caption{
Histogram of parallax for the stars in NGC 2158.
The dashed red line shows all the stars while the 
solid black histogram shows the most probable cluster members.
}
\label{figplxhist}
\end{figure}

\begin{figure}
%\centering
\includegraphics[width=\textwidth]{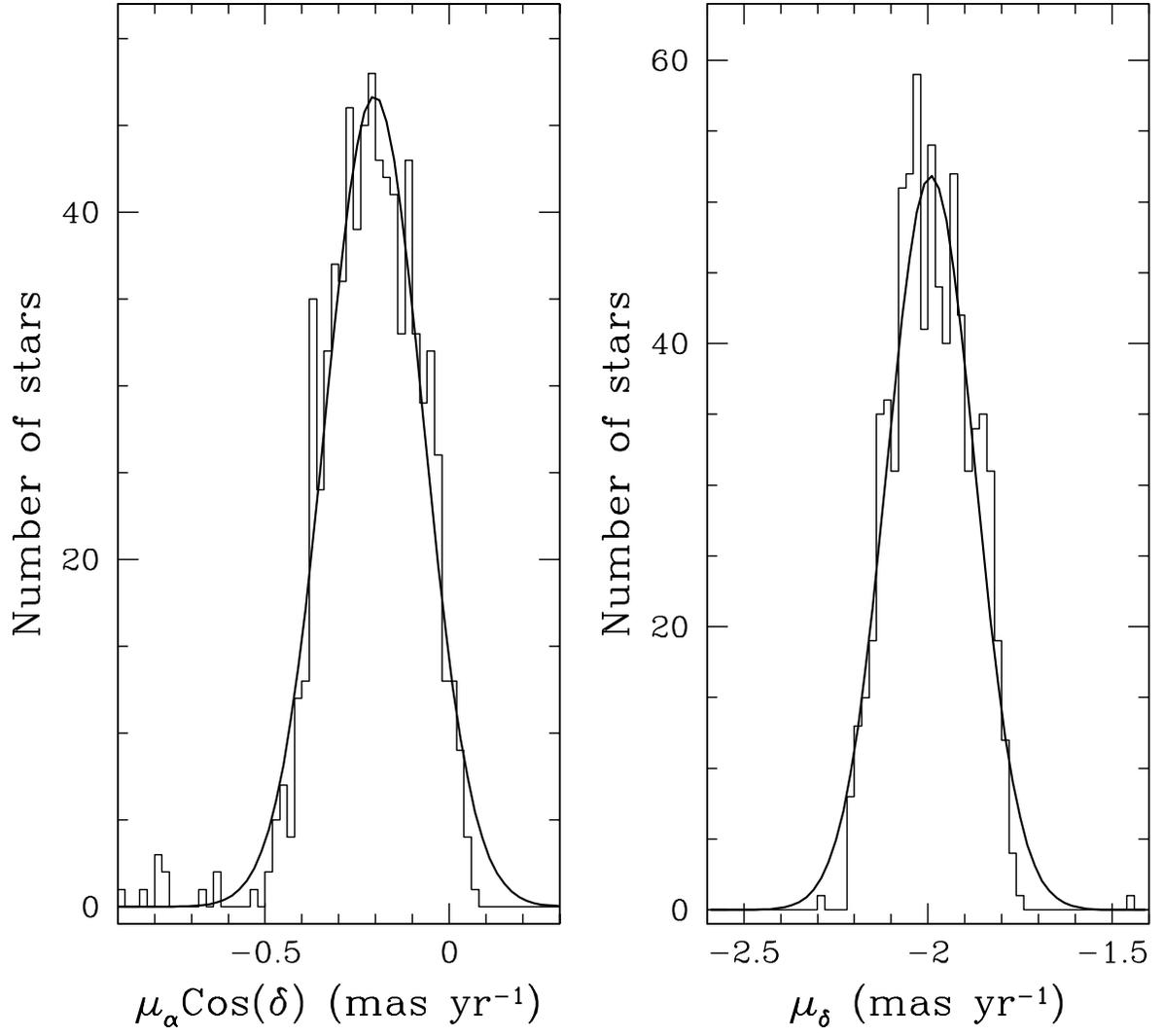}
\caption{
Histogram showing PM distribution and 
the Gaussian fit (black line) to determine the mean value of
PM components. 
}
\label{figpmmean}
\end{figure}

The stars with P$_{\mu} > 90\%$ are shown in a CMD 
plotted in Fig.~\ref{figmp90}. 
We have $\sim$ 800 stars with P$_{\mu} >$ 90\%.
Since the CMD of NGC 2158 does not appear very clean
we considered a high cut off value of P$_{\mu}$($ >$ 90$\%$)  
for the most likely cluster members.
These stars are used to calculate the 
mean parallax and mean cluster PMs. 
The histograms of parallaxes for
all stars in our catalog and the most probable cluster members are shown in Fig.~\ref{figplxhist}.
Figure~\ref{figpmmean} shows Gaussian fitting to the histogram of PM components.
The values of mean PMs for NGC 2158 are 
$(-0.203\pm0.003, -1.99\pm0.004)$ mas~yr$^{-1}$ 
respectively. The estimated values of mean PMs for NGC 2158
are in fair agreement with Cantat-Gaudin et al. (2018).

In this paper, our catalog covers a wider region of NGC 2158 
as compared to the
membership catalog provided by Cantat-Gaudin et al. (2018). 
The Cantat-Gaudin et al. (2018) extracted
the data up to $\sim$10 arcmin radius from the cluster center, 
while our data extraction goes up to a
radius of $\sim$25 arcmin.

\section{Fundamental parameters of the cluster}
\label{FUNDA}

We used the most probable cluster members (P$_{\mu} > 90\%$)
to determine the fundamental parameters of NGC 2158. 

\subsection{Radial density profile}

\begin{figure}
%\centering
\includegraphics[width=\textwidth]{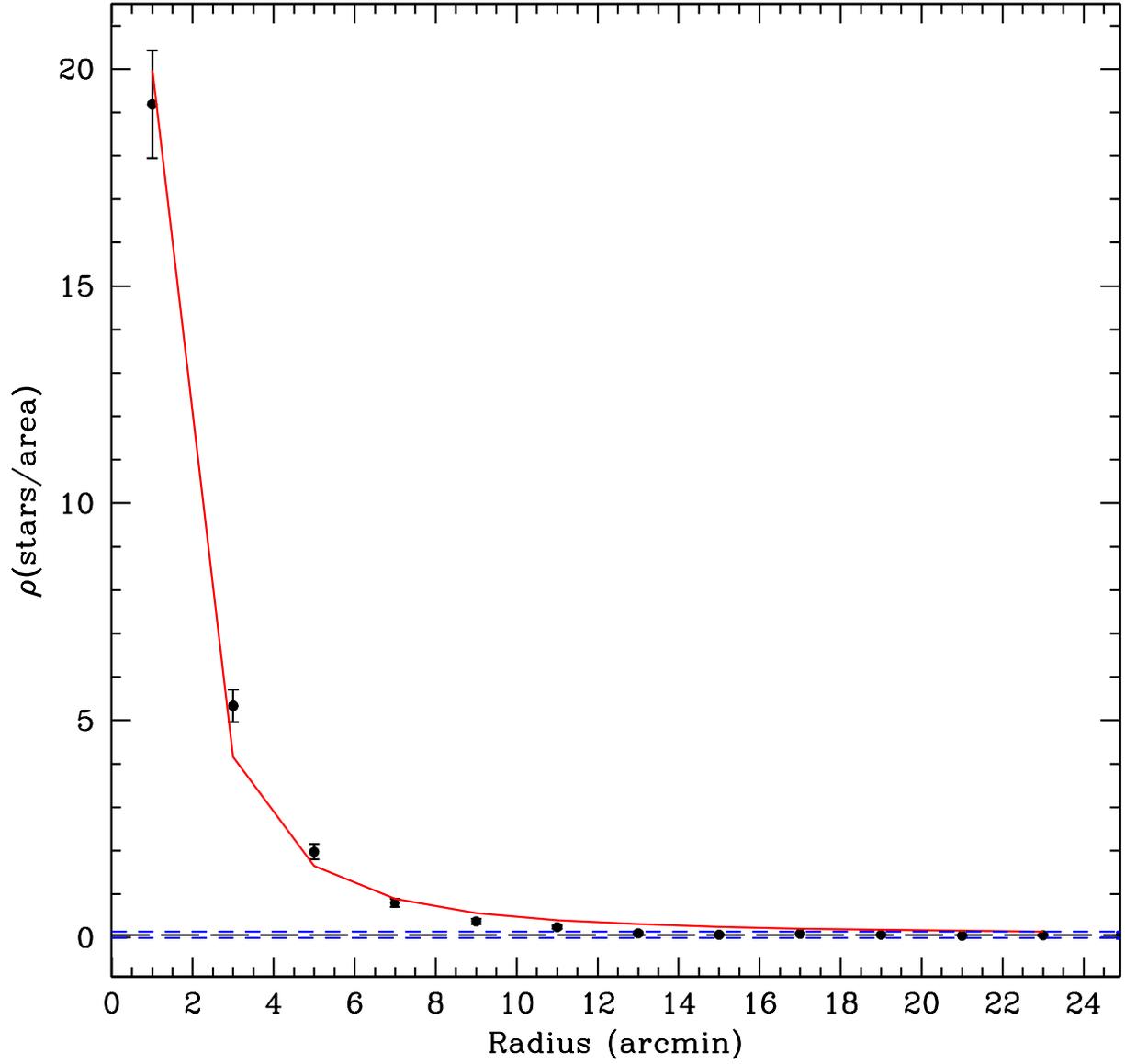}
\caption{
Radial density profile of NGC 2158. The continuous curve here shows the 
King's profile. The background density is shown by horizontal dashed lines
with 3$\sigma$ errors. 
}
\label{fig_rdp}
\end{figure}

The knowledge of a cluster's radius is important to understand the
physical dimensions of the cluster.
We divided the most probable cluster members in concentric radial bins
and determined the number density ($\rho_{i}$ = $\frac{N_{i}}{A_{i}}$)
in each bin. In the $i^{th}$ radial bin, $N_{i}$ is the number of stars
in the corresponding area of the bin $A_{i}$.
The radial density profile (RDP) of NGC 2158 is shown in Fig~\ref{fig_rdp} .
Then, King (1962) profile is fitted on the observed curve of RDP.
The mathematical expression of King profile has been discussed in
Bisht et al. (2020).

For NGC 2158, we obtained the values of 
core radius $r_{c}$, central star density $f_{0}$, 
and background density $f_{bg}$ as: 
$1.04\pm0.21$ arcmin, $38.4\pm12.4$ stars per arcmin$^{2}$, 
and $0.06\pm0.02$ stars per arcmin$^{2}$, respectively.
In Fig.~\ref{fig_rdp}, we plotted two horizontal dotted lines
which show the 3$\sigma$ errors in the background density.
The limiting radius of a star cluster is the limit
after which the stars are no longer gravitationally bound to the cluster.
We have estimated the limiting radius of NGC 2158 as 23.5 arcmin using the
formula mentioned in Bukowiecki et al. (2011).

\subsection{Age and distance}
\label{age}

\begin{figure}
%\centering
\includegraphics[width=\textwidth]{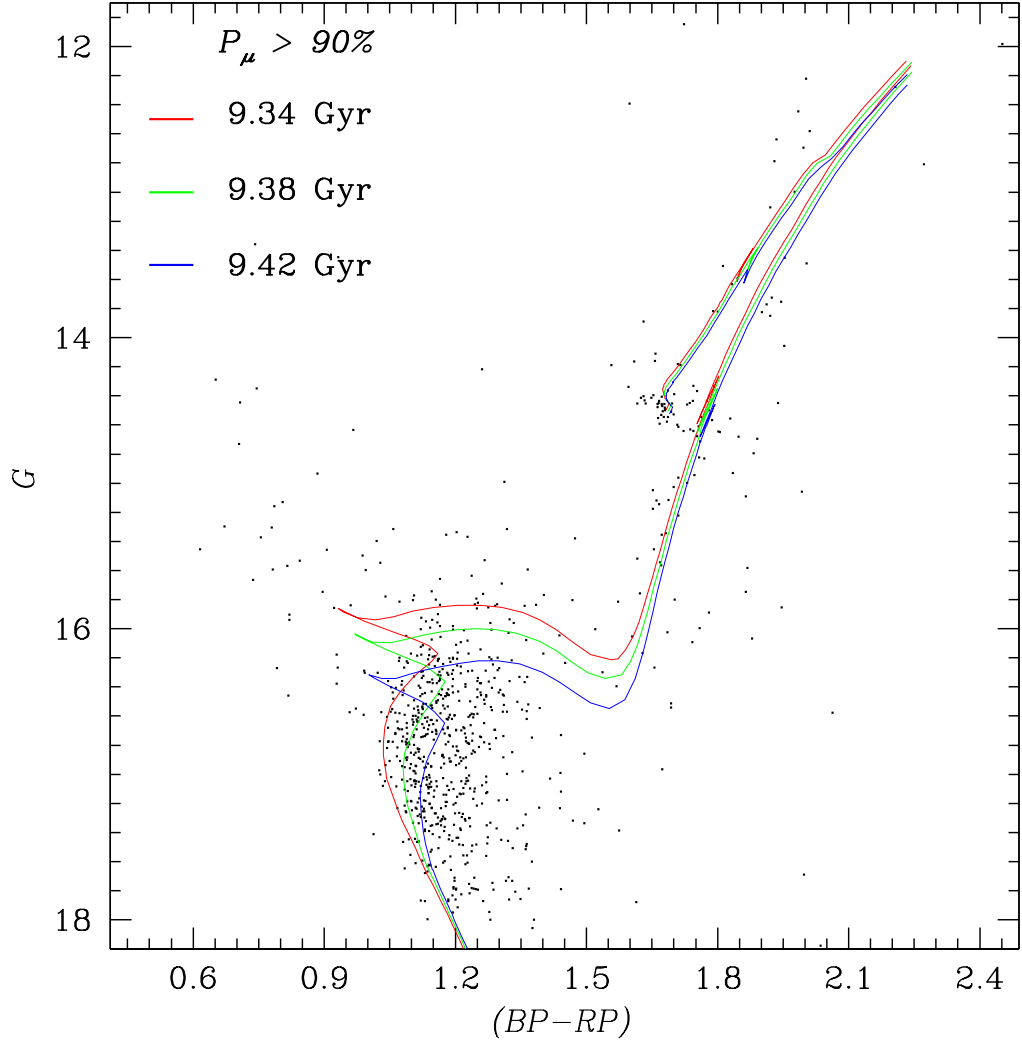}
\caption{Three isochrones of different ages taken from Pastorelli et al. (2019)
is fitted to the $G, (G_{BP}-G_{RP})$ CMD of NGC 2158.
}
\label{figisochrone}
\end{figure}

In order to determine the age, distance and metallicity of NGC 2158,
theoretical isochrones from Pastorelli et al. (2019) are fitted on the 
observational G, $(G_{BP}-G_{RP})$ CMD of the cluster.
We tried various set of isochrones with different age and metallicity values
and finally found the best fitting for the isochrones 
as shown in Fig~\ref{figisochrone} .
The theoretical isochrones of 
different ages (log(age)=9.34, 9.38, 9.42 Gyr) with
metallicity Z=0.004 have been fitted in the CMD based on visual fitting. 
Thus, the best fitted isochrones provides 
log(age) of the cluster as 9.38$\pm$0.04 Gyr which makes NGC 2158 
an old open cluster.
The present value of cluster's age is 
in between the values (2 and 3 Gyr) 
reported by Carraro et al. (2002) and Kharchenko et al. (1997). 
The Gaia-DR2 data also contain $A_G$ values for some stars. 
Using the distance modulus from the isochrone fitting
and median $A_G$ value (0.98) for the most likely cluster members, 
we determined the distance of the cluster as $4.69\pm0.22$ kpc.
This is well in agreement with the distance values 
given by Cantata-Gaudin et el. (2018) and
Kharchenko et al. (2013) written in Section~\ref{INTRO}.

\subsection{Luminosity function and mass function}

\begin{figure}
%\centering
\includegraphics[width=\textwidth]{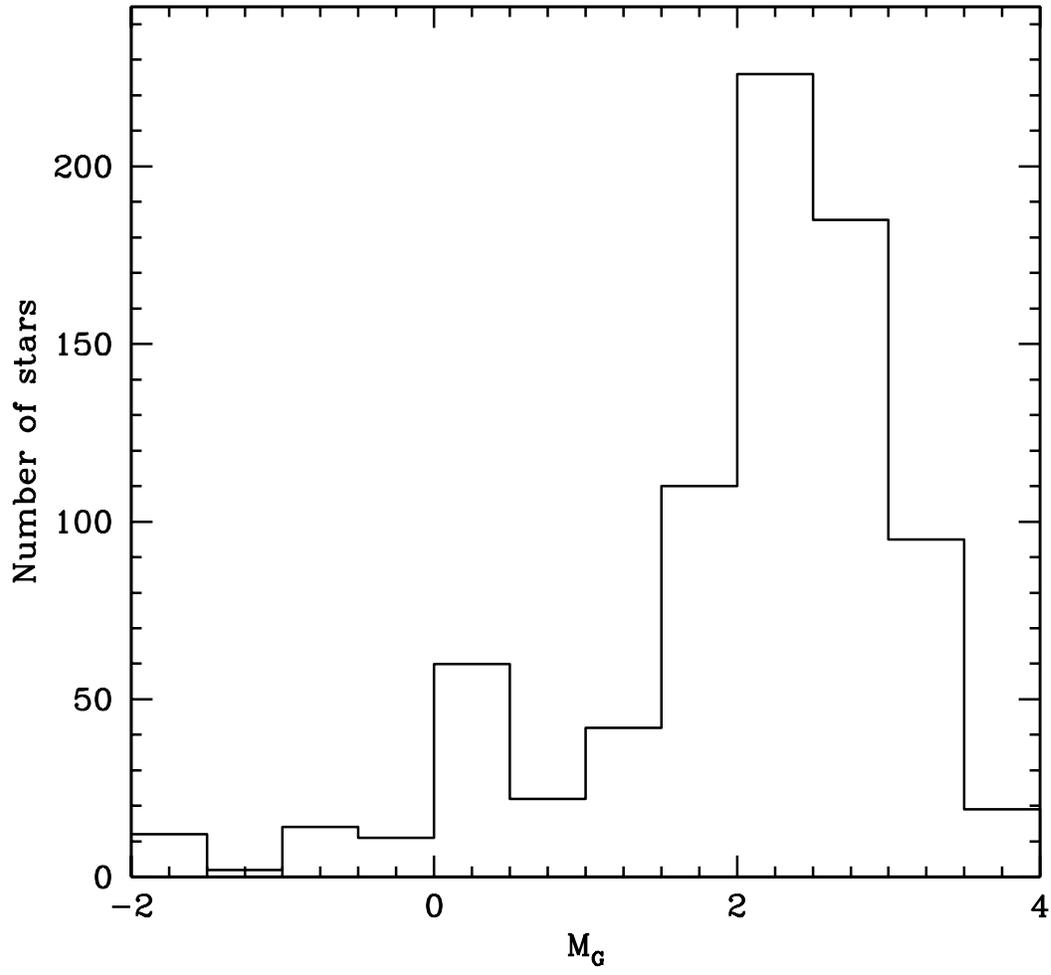}
\caption{
Luminosity function of the most probable members of NGC 2158.
}
\label{figLF}
\end{figure}

The luminosity function (LF) of a star cluster represent the 
distribution of stars according to their brightness.
Before plotting the LF, apparent $G$ magnitudes are first 
converted to the absolute magnitudes using the distance modulus
determined during the isochrone 
fitting to the $G, (G_{BP}-G_{RP})$ CMD 
of NGC 2158. The LF is then plotted as a histogram
with a bin width of 0.5 mag as shown in Fig.~\ref{figLF}.
The LF of NGC 2158 exhibits a rise up to absolute magnitude value of
2.5 mag.

\begin{figure}
%\centering
\includegraphics[width=\textwidth]{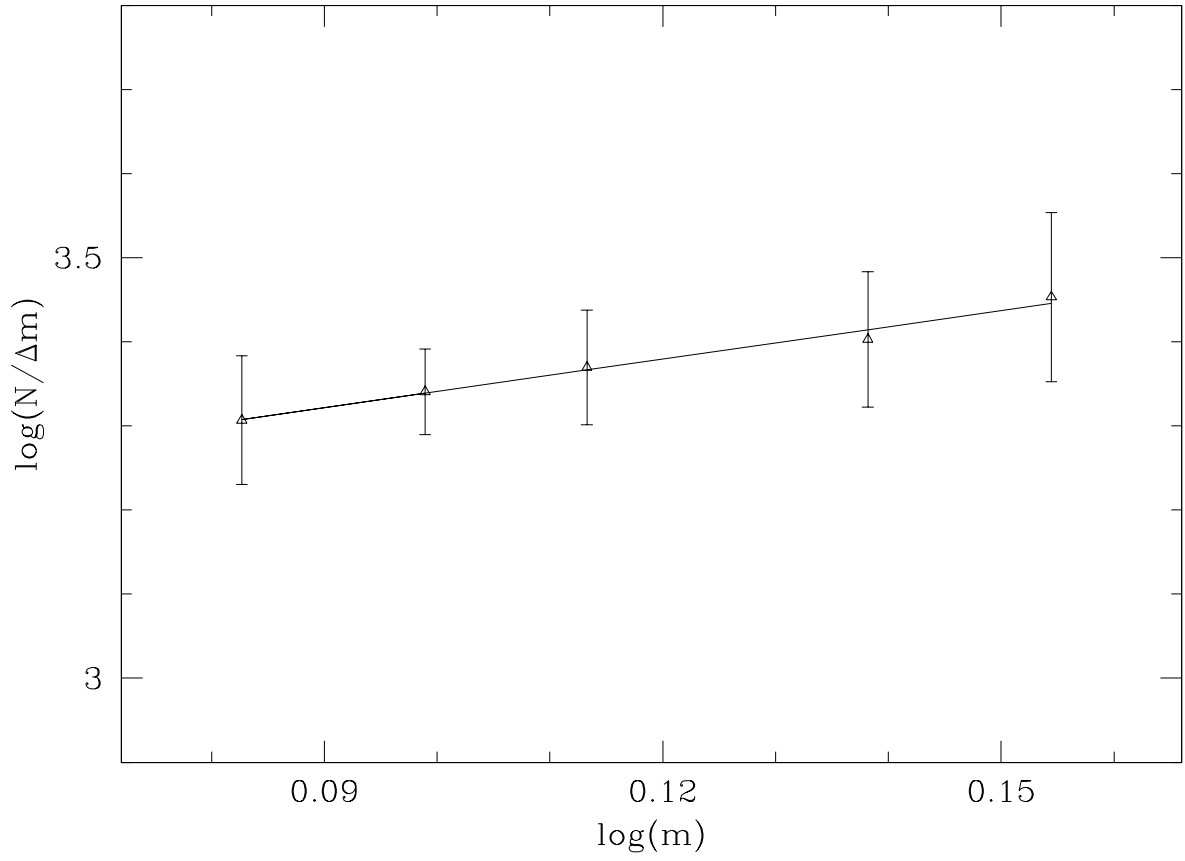}
\caption{
Shown here is the mass function for the cluster NGC 2158. 
The error bars here represent $\frac{1}{\sqrt{N}}$.
}
\label{figIMF}
\end{figure}

Mass function (MF) is directly related to the LF according to the
mass-luminosity relation of stellar astrophysics.
We converted the LF into MF using the theoretical isochrones
given by Pastorelli et al. (2019).
The resulting MF plot is shown in Fig.~\ref{figIMF}.
The MF slope ($x$) can be given according to the following equation:\\

\begin{equation}
\log\frac{dN}{dM}=-(1+x)\log(M)+constant\\
\end{equation}

where $dM$ is a mass range with central mass value of $M$ 
and $dN$ is the number of stars in the mass bin. 
From Fig.~\ref{figIMF}, 
the MF slope for NGC 2158 is obtained as 
$0.93\pm0.14$ within the mass range $\sim$1.17--1.44 $M_{\odot}$.
In this analysis, the magnitude range for MF calculation contains the stars 
in the $G$ magnitude range of 12.5--17.5. 
We did not go deeper in magnitude as Gaia data faces 
an issue of incompleteness towards
the fainter end (Arenou et al. 2018).
The present value of $x$ is flatter than the Salpeter's value (1.35)
which is a signature of dynamical evolution.

\subsection{Mass segregation and relaxation time}

It is argued that mass segregation can also be present from 
the very beginning 
of the cluster, i.e. it could be a signature of cluster formation
(Sagar et al. 1988; Raboud \& Mermilliod 1998; Fischer et al. 1998). 
But, in most cases, it is 
found that the interaction among the stars in a cluster leads to the energy exchange
(Inagaki \& Saslaw 1985; Baumgardt \& Makino 2003).
As a result, the more massive stars move closer to the center to make the 
kinetic energy equipartition possible. Thus, the velocity distribution in a cluster
becomes almost Maxwellian. The time taken for this dynamical evolution is 
given by the relaxation time. The following equation 
given by Spitzer \& Hart (1971) defines the relaxation time:\\

\begin{equation}
\hspace{2.0cm}T_{E} = \frac {8.9 \times 10^{5} N^{1/2} R_{h}^{3/2}}{ <m>^{1/2}log(0.4N)}
\end{equation}

where $N$ is the number of most probable members of NGC 2158.
$R_{h}$ is the half-mass radius of the cluster
and  $<m>$ is the average mass of the stars in the unit of $\sim M_{\odot}$.
The relaxation time for NGC 2158 is calculated as 
$584.98\pm5.63$ Myr.
This is clearly less than the cluster's age calculated in Section~\ref{age}.
Ultimately, it can be concluded that NGC~2158 is dynamically relaxed. 

\section{Determination of the cluster apex}
\label{adapex}

To highlight stellar groups with a common motion in space,
we adopted the AD diagrams method.
This method uses the concept of ``individual star apex''.
This term was introduced by analogy with the apex of the
Sun or the apex of the cluster.
Apex is a point on the celestial sphere
towards which an object moves; in this case, a star or a star cluster.
Individual apexes can be obtained by placing the beginning of
a star’s spatial velocity vector to the point of observation
and extending it to the intersection with the surface of the celestial sphere.
The intersection point will be the star apex.
In other words, we mean an individual apex of a star
as a point on the celestial sphere with coordinates $(A, D)$
in the equatorial coordinate system,
which is directed by a vector parallel to the star’s spatial velocity vector,
plotted from the observation point.
The AD-diagram represents the positions of individual apexes
in equatorial coordinates.

We considered the apexes of cluster stars and determined the average value.
The proximity of the points in the $AD$ diagram
indicates the parallelism of the corresponding spatial velocity vectors.
According to the concentration of points in the diagram,
it is possible to distinguish groups of stars with a common motion in space.

This method is convenient for its simplicity and clarity. 
%Unlike UVW diagrams,
%where it is necessary to consider an ellipsoid of velocities,
%the stellar apex method selects co-directional vectors in the plane,
%and there are no restrictions on the velocity vector modulus.
Description of the method, the charting technique,
and formulas for determining error ellipses
can be found in Chupina et al (2001).

To calculate the individual star apex, we need distance, proper motion, 
and radial velocity (RV). 
We detected 12 stars in Gaia DR2 release, 
for which we could find all the kinematical parameters 
including the RV measurements from Gaia-DR2. 
To expand the sample of stars, we cross-matched our most 
probable cluster member stars with 
The Large Sky Area Multi-Object Fibre Spectroscopic Telescope (LAMOST) DR5 catalog
(Luo et al. 2019).
For LAMOST stars with more than one RV measurements,
we accepted the value with the less relative error. 
Thus, we received 13 additional stars from the LAMOST survey with available RVs.
In the resulting list, we have 25 stars with a complete set of astrometric data 
and distances of stars to the Sun (13 from LAMOST DR5 and 12 Gaia DR2). 
In order to check whether there is a systematic difference between 
the RV data from LAMOST and Gaia DR2, we compared the RV measurements 
for more than 8000 stars within a radius of 5 degrees 
around the center of NGC 2158. 
We detected a systematic difference of $\sim5$ km s$^{-1}$ between the two RV data. 
This value is supported by a similar study conducted by Du et al. (2018) 
for a different sample of stars than ours. 
We applied this shift to the RVs of 13 stars 
taken from LAMOST to bring the RV values to the same level as Gaia-DR2. 
The correction is applied in the sense
that for the LAMOST data, $V_r$=$V_r^{LAMOST}$+5~km s$^{-1}$.
The corrected values are listed in Table~\ref{tabapex}.

In Table~\ref{tabapex}, there is a scatter in the values of RV.
Considering the values of $\sigma_{V_r}$ in Table~\ref{tabapex},
the measurement errors might not be contributing to this scatter. 
The mean value of errors in RVs is $\sim$2 km s$^{-1}$ for Gaia-DR2 stars 
and about 5.6 km s$^{-1}$ for the LAMOST data. 
The spread in the RVs exceeds these mean errors. 
Since this RV spread is present for both Gaia-DR2 or LAMOST data, 
it might not be caused by 
the measurement method and its overall precision. 
One possible reason behind this scatter could be the fact that 
the real measurement errors are much higher than the formal errors. 
There is also a possibility that some stars may be parts of the 
binary or other multiple systems, which could have 
affected the measurements of their RVs.

Please note that we used the stellar distances from Bailer-Jones et. al (2018). 
The data is presented in Table~\ref{tabapex}. 
The columns of Table~\ref{tabapex} contain:
the star ID number according to our catalog, 
the membership probability ($P_{\mu}$), 
equatorial coordinates ($RA$, $DEC$), 
coordinates of the apex ($A$, $D$), 
distance and its lower and upper bounds taken from Bailer-Jones et. al (2018), 
RV and its error ($V_r$, $\sigma_{V_r}$), 
and PM components ($\mu_\alpha cos\delta$, $\mu_\delta$), 
with their errors ($\sigma_{\mu_\alpha cos\delta}$, $\sigma_{\mu_\delta}$). 

\begin{figure*}
\centering
\includegraphics[width=7cm]{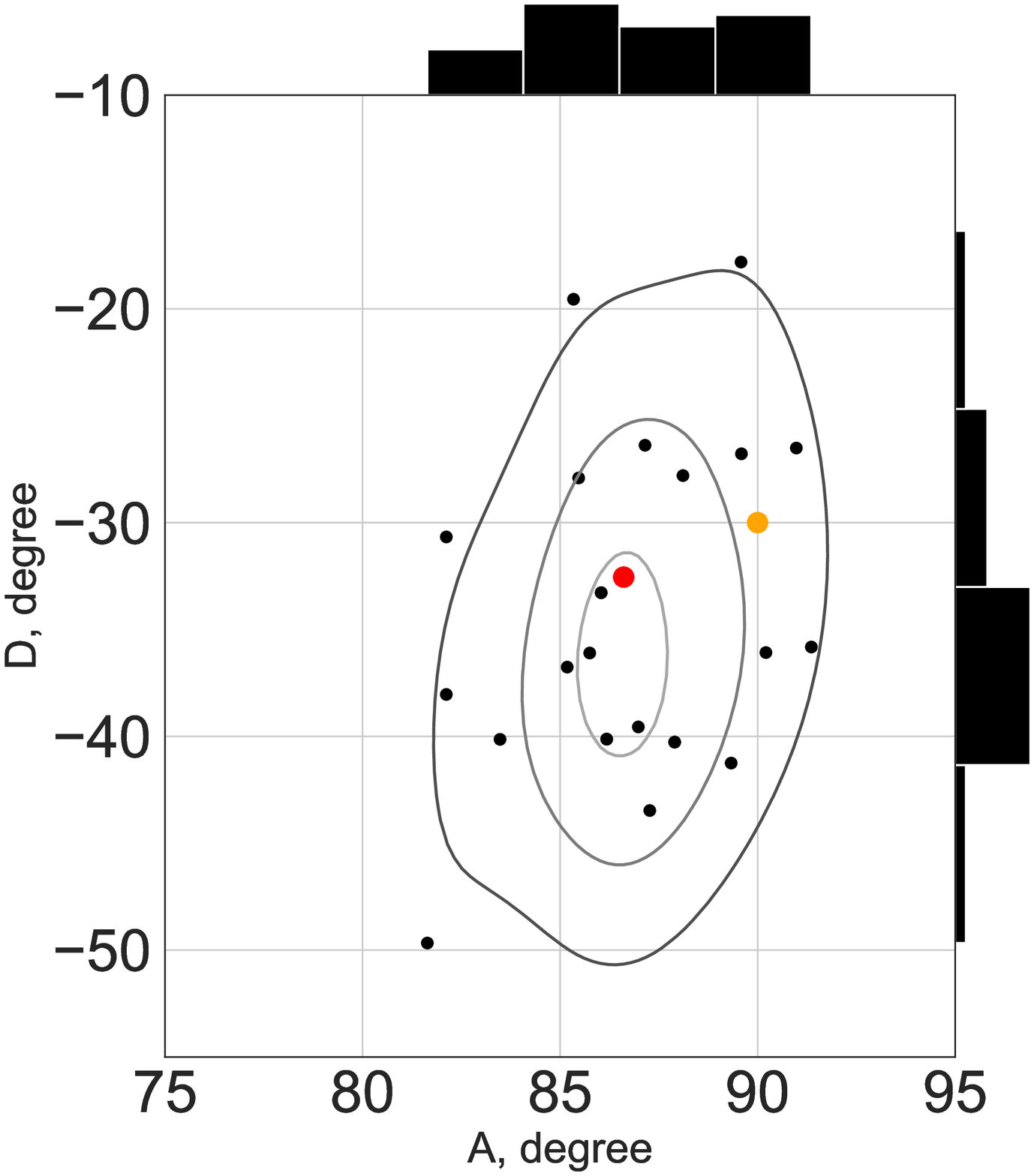}
\includegraphics[width=6.6cm]{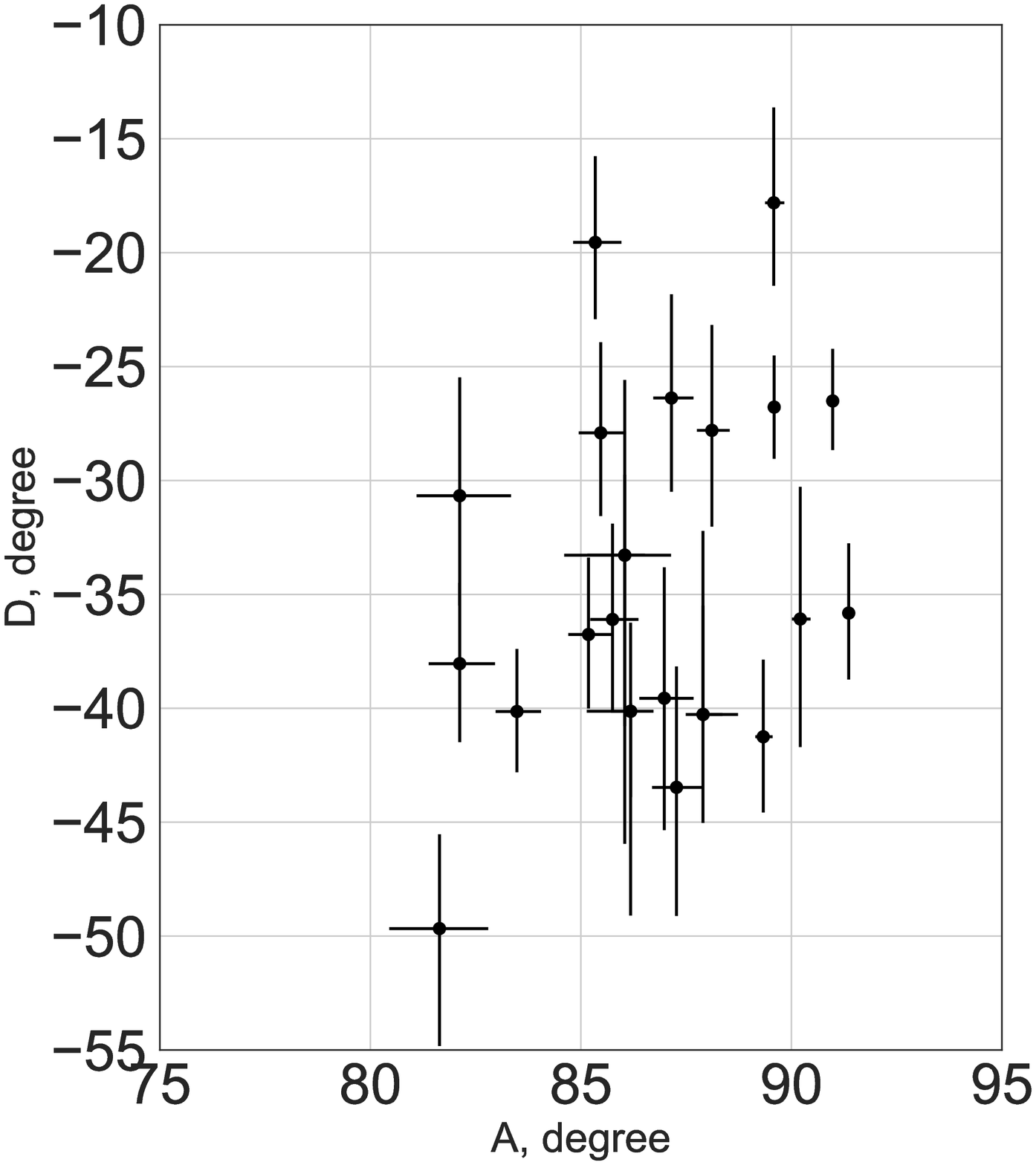}
\caption{Apex diagram for the NGC~2158 cluster.
In the left panel the 
orange dot shows the position of the Solar antapex ($A=89^\circ$, $D=-30^\circ$; Cox 2000). 
The red dot represents cluster apex position 
$A$=87.24$^\circ\pm$1.60$^\circ$, $D$=$-$36.61$^\circ\pm$5.30$^\circ$.
The right panel shows errors in $(A, D)$.}
\label{figAD}
\end{figure*}

\begin{figure*}
\centering
\includegraphics[width=16cm]{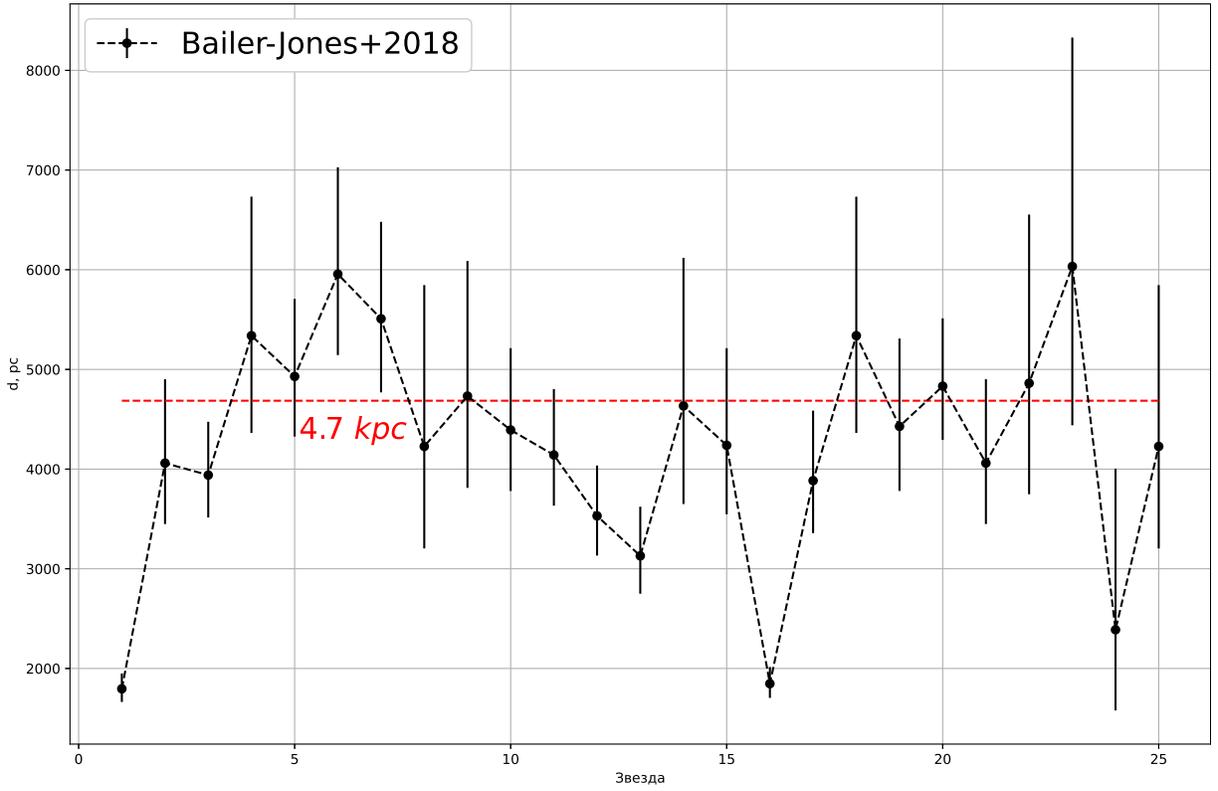}
\caption{Bailer-Jones et. al (2018) distances with error bars 
(see table~\ref{tabapex}). 
The red dashed line shows the determined by us distance for NGC~2158.}
\label{figderr}
\end{figure*}

The results of determining the apexes are shown in Fig.~\ref{figAD}.
The contours in Fig.~\ref{figAD} show the density of distribution of 
points on the AD-diagram. 
They emphasize the uneven distribution of the star apexes.
The histograms along the coordinate axes show the same thing.
The distribution are uneven and there are maxima.
There are significant differences in the uncertainty of 
determining the distances to the stars,
as can be seen from the lower and upper bounds of the distance in Table~\ref{tabapex}, 
These differences are illustrated in Fig.~\ref{figderr}, 
where for each star we have plotted spread in the distance value 
according to data of Bailer-Jones et. al (2018). 
Using Table~\ref{tabapex}, the length of lower $l_{b}$ and upper $l_B$ bars 
are determined as: 
$l_b = Distance (pc) - Low~bound(pc)$, $l_B = Upper~bound(pc) - Distance (pc)$. 
In order to take these uncertainties into account, 
we used the following averaging procedure considering weights for each star. 
Due to weighty nature of the relative errors, 
contribution of each star was calculated assuming its weight: 
$p = \frac{1}{|D_{NGC} - D_{BJ}|}$, 
where $D_{NGC}$=4.69~kpc (Section~\ref{age}) and $D_{BJ}$ 
is the average distance (see distance column in Table~\ref{tabapex}). 
Conclusively, we determined the cluster apex 
$A$=87.24$^\circ\pm$1.60$^\circ$, $D$=$-$36.61$^\circ\pm$5.30$^\circ$.

Figure~\ref{figAD} shows the Solar antapex with coordinates 
$(A, D)$ = (89$^\circ$, $-$30$^\circ$ Cox 2000). 
As we see in Fig.~\ref{figAD}, the movement of the Sun and NGC 2158 
occur in opposite directions. 
We confirm it further in Section~\ref{orbit} with kinematic data. 
In Fig.~\ref{figAD}, a separate panel shows 
the individual uncertainty in the apex positions 
due to the uncertainty in determining the distances 
according to Fig.~\ref{figderr}. We see that errors are significant, 
but not critical for such a distant cluster, 
where higher accuracy cannot be expected.

The value of the cluster apex will be used
for determining ($\mu_{U}$,$\mu_{T}$) diagram
presented in Section~\ref{secmuut}.

\subsection{Diagram of $\mu_{U}$ and $\mu_{T}$}
\label{secmuut}

In contrast to RVs, PMs are usually available for
a much larger number of stars. 
We have 25 most likely cluster member stars with 
available RVs and 800 most probable members with PMs. 
It will also demonstrate how legitimate it is to consider the 
apex coordinates obtained in Section~\ref{adapex} 
as acceptable for assessing the direction of motion of NGC~2158 in space.

The coordinate system of $\mu_U$-$\mu_T$ diagram is selected as follows:
the reference axis $\mu_{U}$ is directed to a point on the celestial sphere
representing the apex position,
and $\mu_{T}$ axis is directed perpendicular to the $\mu_{U}$ axis.
Thus, on the ($\mu_{U}$, $\mu_{T}$) diagram,
apex direction will be determined by stars with $\mu_{T} \approx 0$.

This diagram is constructed by calculating the 
$\mu_{U}$, $\mu_{T}$ values, 
determined in new coordinate system with $\mu_{T}$ directed to the apex. 
Using equatorial apex coordinates obtained in Section~\ref{adapex},
we constructed a diagram in the ($\mu_{U}$, $\mu_{T}$) coordinates for NGC 2158.

\begin{figure*}
\centering
\includegraphics[width=16cm]{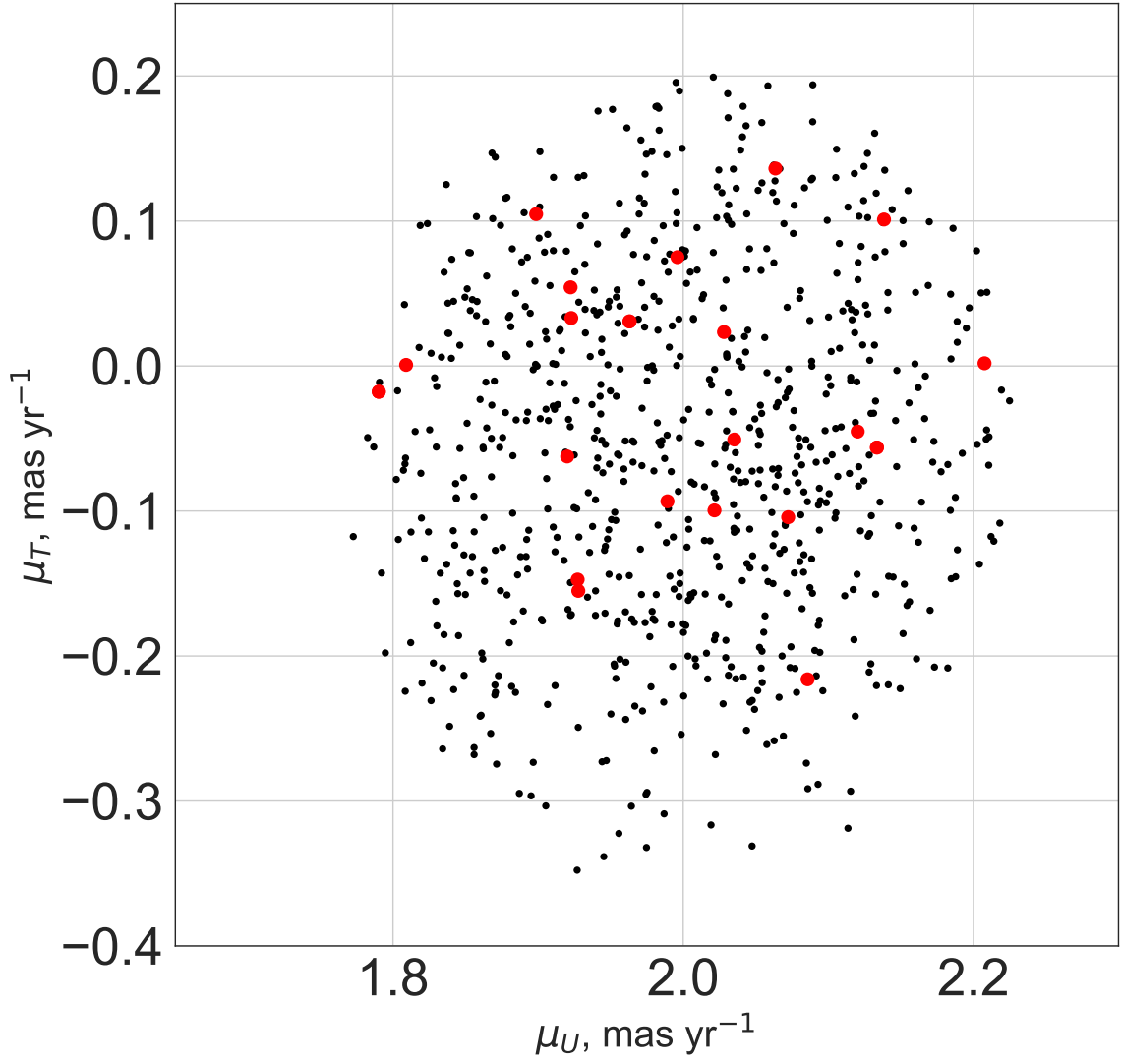}
\caption{The $\mu _{U}$-$\mu _{T}$ diagram for 
800 most probable member stars. The red dots represent the 
stars from table~\ref{tabapex}.}
\label{figmuut}
\end{figure*}

The resulting diagram in $\mu_{U}$, $\mu_{T}$ coordinate system is 
presented in Fig.~\ref{figmuut}. 
The figure shows both sets of data: 25 stars with 
available RVs (see Table~\ref{tabapex}) 
and 800 stars from our membership catalog with membership probability $>90\%$. 
In general, the distribution of points in Fig.~\ref{figmuut} 
is uniform for both sets. 
In order to numerically estimate the reliability of the obtained values 
of the cluster apex, we calculated the average value of 
$\mu_{T}$ for 800 stars, which is $\mu_{T}$ = $-$0.066. 
This value being close to zero 
indicates a good quality of our definition of the average cluster apex.

\section{Orbit of NGC 2158}
\label{orbit}

We have performed orbital integrations using 
``MWPotential2014'', the default galpy potential of the Milky Way (Bovy 2015). 
This potential includes a power law with cut-off bulge, 
a Miyamoto-Nagai disk (Miyamato \& Nagai 1975) 
and a Navarro-Frenk-White halo (Navarro et al. 1995). 
We have set the distance to the Galactic center as $R_0=8.178~kpc$ 
and the circular velocity at the Sun's radius as $V_0=232.8~km~s^{-1}$. 

\begin{figure*}
\centering
\includegraphics[width=10cm]{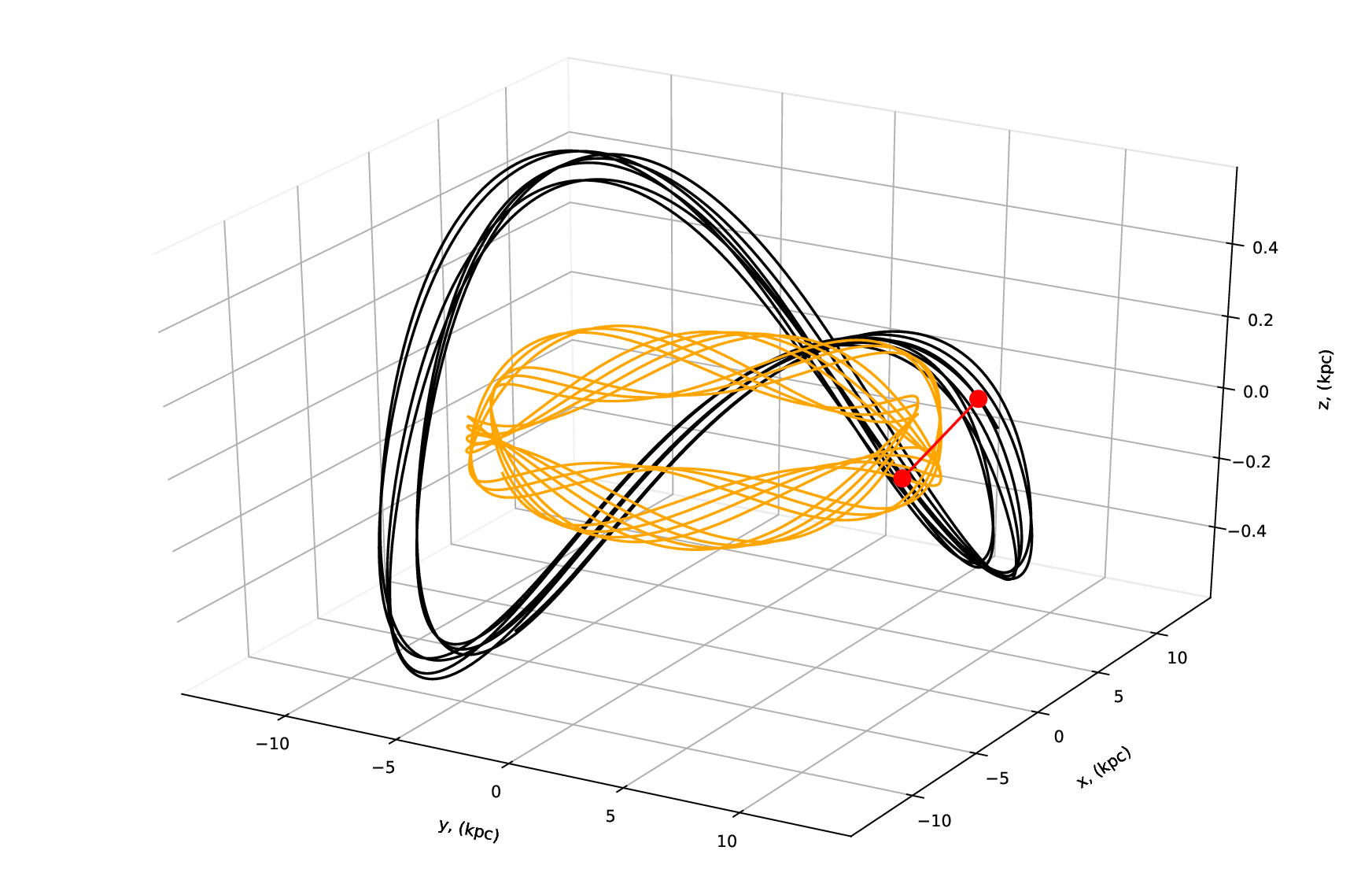}
\includegraphics[width=6cm]{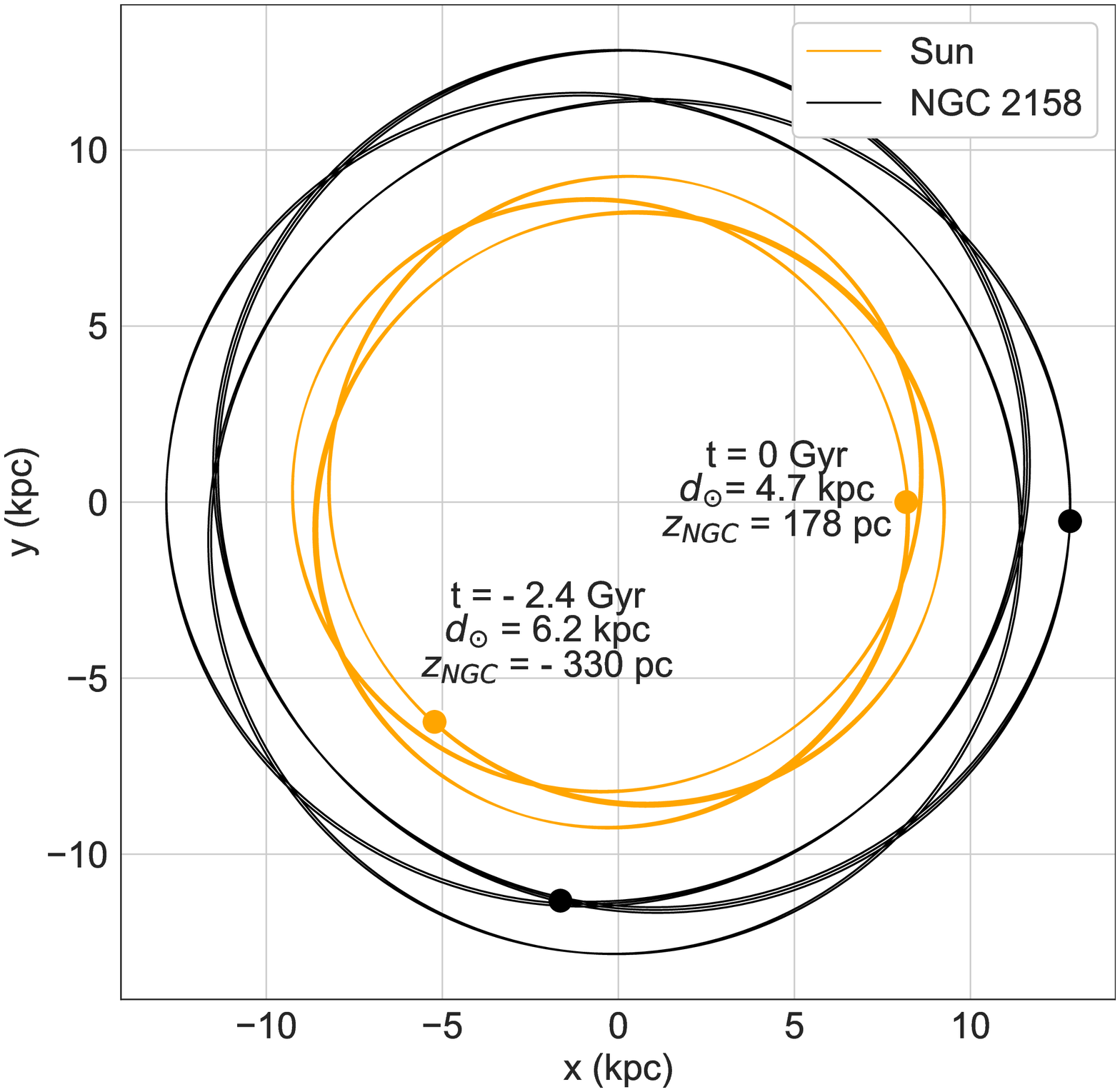}
\caption{The cluster's orbit in 3D (left panel)
and orbit in the XY Galaxy plane (right panel) 
integrated backward to 2.4 Gyr.
Dots in the right panel indicate positions of the Sun and cluster at 
the time indicated in the labels. 
In the left panel, the red line represents the minimum distance 
(d=2.7 kpc at t=$-$2.3 Gyr) between Sun and NGC 2158. 
The label gives information about the $Z$-coordinate
at the indicated times and the cluster's distance from the Sun (d) 
for t=0 and t=$-2.4$~Gyr.
}
\label{figorbit}
\end{figure*}

Figure~\ref{figorbit} shows the integrated orbits of 
Sun and NGC 2158 in Cartesian Galactocentric coordinate system 
backward in time according to cluster age (2.4 Gyr) 
determined in this paper. 
As an input, we used cluster parameters presented in 
Table~\ref{taborbitinput}: PMs ($\mu_{\alpha}cos\delta$, $\mu_{\delta}$), 
distance from the Sun $d$, right ascension $RA$ and declination $DEC$, 
and RV $V_{r}$ 
which was calculated as average from 12 Gaia DR2 stars in Table~\ref{tabapex}.
We avoided the RV values from LAMOST DR5 here due to their 
larger errors which would have propagated to cause larger errors in our output 
orbital parameters. 

The left panel of Fig.~\ref{figorbit} shows 3D orbits. 
According to the Z-coordinate, the cluster oscillates
in about every 250 million years, rising above the plane of the 
disk up to 529 pc. 
We could roughly estimate the birthplace of 
cluster at its zero age at Z = $-$330 pc. 
The current position of the cluster is at Z = 178 pc.
It is likely that the cluster could have been born 
farther from the plane of the disk as compared to its current position.

As we can see in Fig.~\ref{figorbit} (right panel) that 
during its lifetime, the cluster and Sun move in an 
approximately circular orbits (in projection onto the Galactic plane) 
around the Galactic center. 
During its lifetime, 
NGC~2158 have had more than 10 revolutions around the Galactic center.

The errors in PMs and radial velocity can affect 
the cluster's orbital parameters as well as 
the positions of the cluster and the Sun in space. 
From Section~\ref{MP}, we have the PM values with errors
($\mu_{x}=-0.203 \pm 0.003$, $\mu_{y}=-1.99 \pm 0.004$ mas yr$^{-1}$).
Using the standard deviation of 12 Gaia-DR2 stars from Table~\ref{tabapex} 
to calculate error in radial velocity, 
we have the radial velocity value with error ($25.1 \pm 5$ km s $^{-1}$). 
The resulting orbital parameters after considering these 
errors are as follows: 
apocentric radii = $12.8 \pm 0.03$~kpc, 
pericentric radii = $11.1 \pm 0.05$~kpc, 
eccentricity $e= 0.07 \pm 0.003$, 
$Z_{max}=529\pm7$~pc, 
($U,V,W$) = ($-$21.5 $\pm$1.7, $-$39.2 $\pm$1.7, $-$24.6 $\pm$1.7) km s$^{-1}$
and orbital period $T_r \sim 0.2$~Gyr.
The output results are also presented in the Table~\ref{taborbitinput}. 

\begin{figure*}
\centering
\includegraphics[width=8cm]{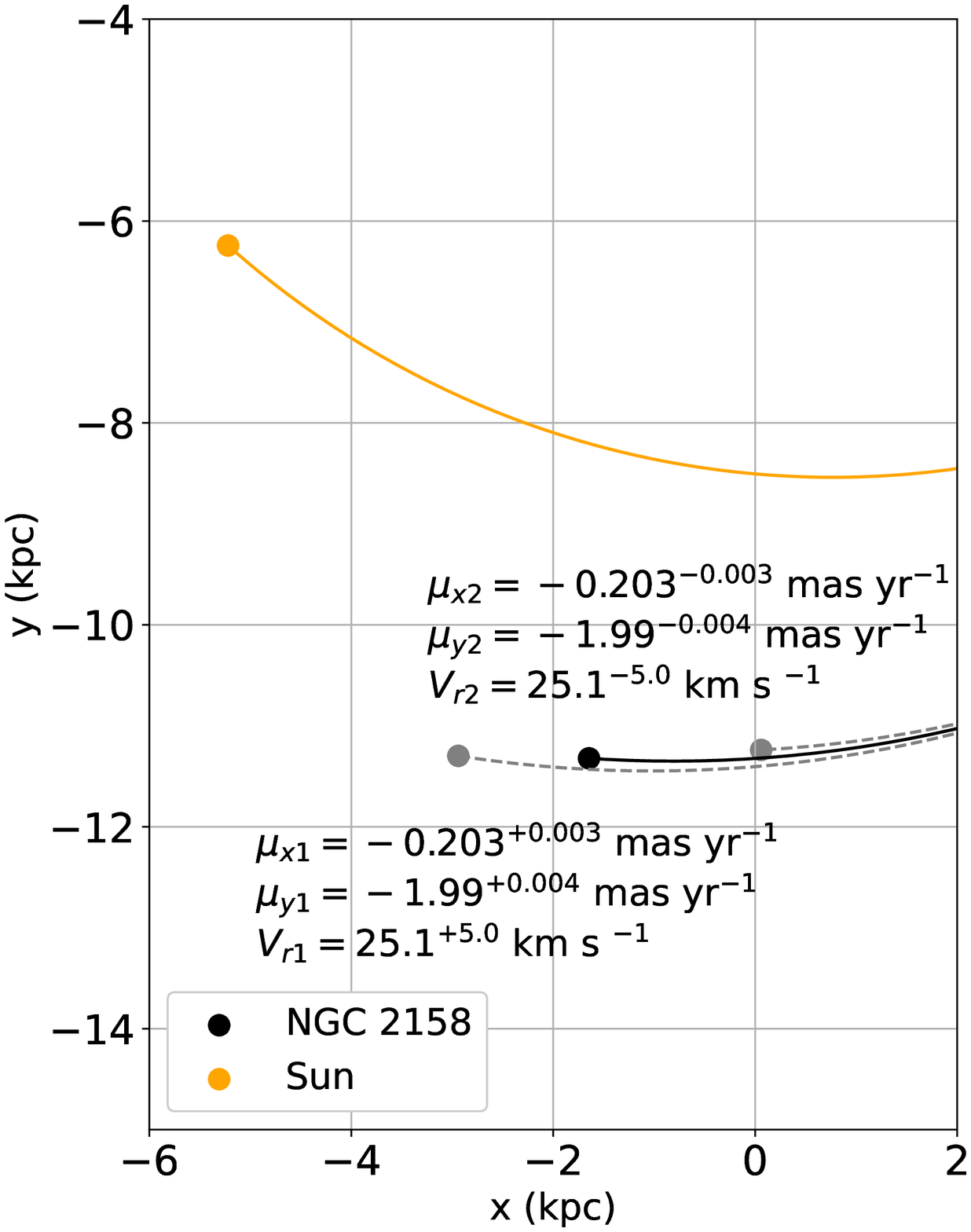}
\includegraphics[width=7.15cm]{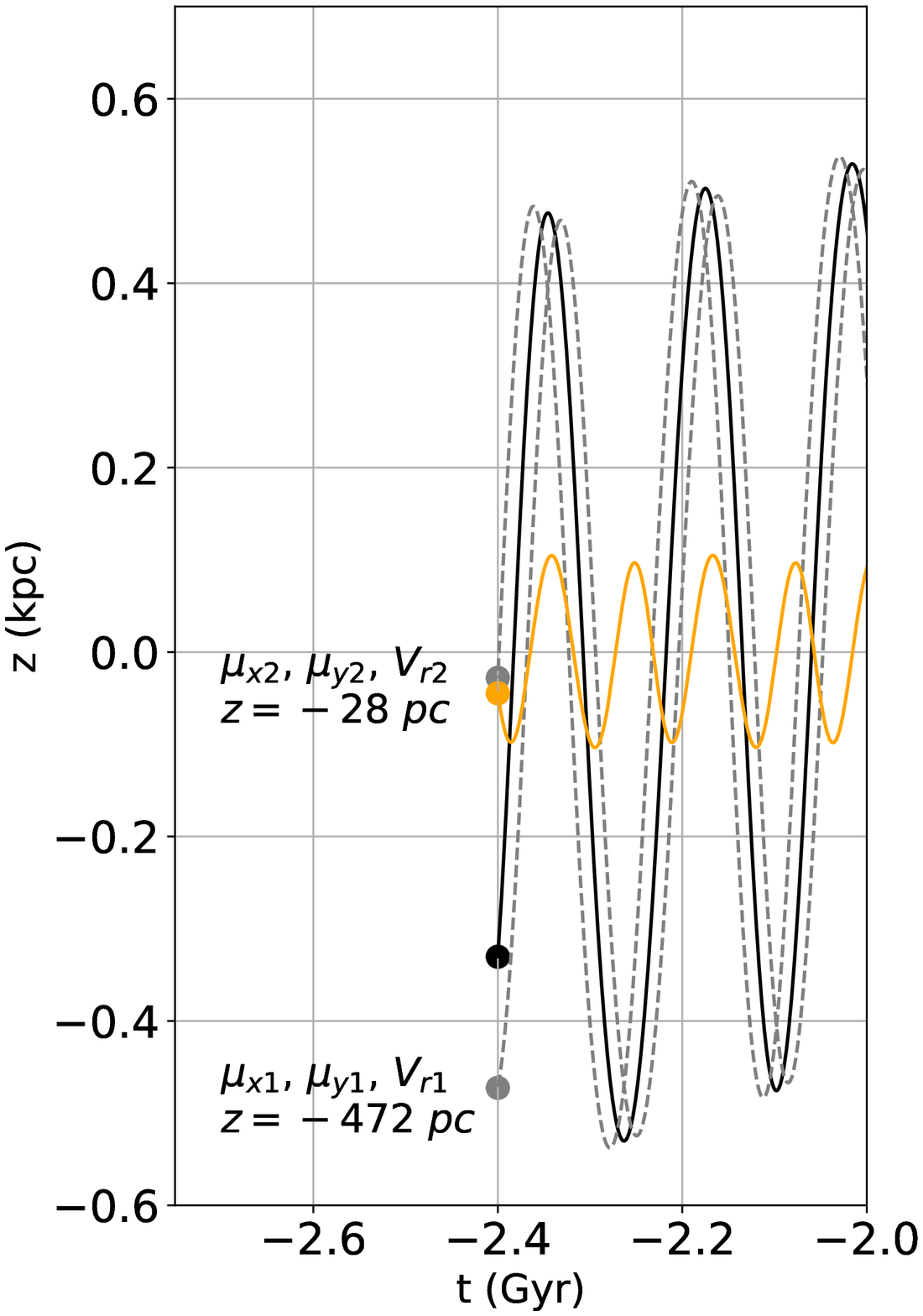}
\caption{The cluster's orbit integrated with different input parameters
coming from the inclusion of errors in PMs and RV
at the time of possible birth $t=-2.4$ Gyr. 
Orbits are plotted in Galactocentric rectangular coordinates in 
$XY$ plane (left panel) 
and $Z$-coordinate versus time (right panel).
The input parameters are also mentioned in the first panel of the figure.
Black and orange orbits represent NGC 2158 and the Sun respectively 
(same notation as Fig.~\ref{figorbit}). 
}
\label{figorbit_e}
\end{figure*}

The effect of PMs and radial velocity errors on the integrated orbits is shown in Fig.~\ref{figorbit_e}. 
As can be seen, despite starting from a single place in the Galaxy, 
different input parameters lead to different birth locations of the cluster. 
In the $XY$ plane of Fig.~\ref{figorbit_e}, the possible place of birth of the cluster
has an uncertainty of $\sim$2~kpc. 
While it looks significant considering the value,  it is not important 
as the cluster stays at the same Galactocentric distance. 
From the $Z$-coordinate (the right panel of Fig.~\ref{figorbit_e}), 
we can see that the cluster could have formed 
in the southern hemisphere of the Galaxy.
However, from the simulations, it is
difficult to assess how far from the Galactic plane the cluster was born.

\subsection{Discussion on the origin of the cluster}
\label{origin}

NGC 2158 is an old cluster located at the Galaxy's periphery.
In this context, it is intriguing to discuss several possibilities for the origin of NGC 2158.
The possible scenarios are discussed as follows\

\begin{enumerate}

\item The location of the cluster in the outer disk opens 
to the possibility that the cluster was
accreted. However, our findings do not support this idea. 
We find that NGC 2158 kinematics is consistent with a disk orbit.
Backward integration (see Section \ref{orbit}) 
shows the possible birthplace of the cluster at
$X=-1.5\pm1.4$ kpc, $Y=-11.3\pm0.1$ kpc, $Z=-0.25\pm0.2$ kpc.\

\item Anders et al. (2017) discuss that the radial mixing (by migration and heating) 
can explain the presence of intermediate-age, high-metallicity open clusters 
in the solar neighborhood. 
We discuss the possibility that NGC 2158 location is due to migration.
NGC~2158, in accordance with the parameters presented in Section~\ref{orbit}, 
is located at $R_{Gal} < 12$ kpc, $Z < 0.529$ kpc. 
From isochrone fitting, the log(age) value of NGC 2158 is 9.54 Gyr. 
The cluster's metallicity value Z= 0.004, can be transformed into $[Fe/H] = -0.68$.
The metallicity value obtained by us 
corresponds approximately to the position of the cluster in the Galaxy
when compared with Anders et al. (2017) data compilation. 
Our metallicity value is also in agreement with the value ($[Fe/H] = -0.60$) obtained by
Carraro et al. (2002).

\end{enumerate}

\section{Conclusions}
\label{CON}

In  this paper, we have done photometric and kinematical analysis of 
old open cluster NGC 2158 using Gaia DR2 data. We
identified 800 member stars with membership probability higher than $90\%$. 
The main results of the present investigation
are as following:

\begin{itemize}

\item The mean PMs of the cluster are estimated as 
$-0.203\pm0.003$ and $-1.99\pm0.004$ mas yr$^{-1}$ 
in both the RA and DEC directions respectively.\

\item  Distance to the cluster NGC 2158 is estimated as $4.69\pm0.22$ kpc. 
The log(age) is calculated as $9.38\pm0.04$ Gyr by comparing
the cluster CMD with the theoretical isochrones given by 
Pastorelli et al. (2019). \

\item  The mass function slope is determined as $0.93\pm0.14$. 
The obtained slope is quite flat, which is a hint of 
dynamical evolution in the region of NGC 2158.\
       
\item  Mass segregation has been studied for NGC 2158. 
The dynamical relaxation time is less than the cluster's age, which indicates
that this object is dynamically relaxed.\

\item The apex coordinates of NGC 2158 are determined using $AD$-diagram method.
Further, the apex and kinematics is analyzed using $\mu_{U}$ -- $\mu_{T}$ diagram. 
The estimated cluster's apex is
$A$=87.24$^\circ\pm$1.60$^\circ$, $D$=$-$36.61$^\circ\pm$5.30$^\circ$.

\item The orbit of the cluster in the Galactic potential is studied.
We find that the cluster follows a disk orbit.
The scenarios we considered included the current position as well as 
the cluster's motion in the past.
The spatial velocity of the cluster is equal to 51~km s$^{-1}$ 
relative to the Sun. 
Our model shows a minimal distance of 
$2.7 \pm 0.75$ kpc between the orbits of the Sun and the cluster 
(left panel of Fig.~\ref{figorbit}). 
Figure~\ref{figorbit_e} shows the place (with position uncertainties) 
in the Galactic disk where NGC 2158 could have been formed 
2.4 billion years ago. 
However, if we take into account the errors on the
age of the cluster and on its PMs,
then in Fig.~\ref{figorbit_e} (right panel), 
a significant oscillation along the $Z$ coordinate up to 529
pc is detected.		

\item Various scenarios are discussed to explain the origin 
and the position of the cluster in the outer disk, 
namely accretion and migration. 
NGC 2158 orbit is consistent with disk kinematics, 
leading us to discard the possibility that the cluster is accreted. 
Its metallicity is not inconsistent with the position in the disk. 
This gives no support to the possibility that
the current location of the cluster in the disk is due to migration.

\end{itemize}

\section*{Acknowledgment}
We are thankful to the referee for the constructive comments 
that significantly improved this paper.
This work is supported by the grant from the 
Ministry of Science and Technology (MOST), Taiwan. 
The grant numbers are MOST 105-2119-M-007 -029 -MY3 
and MOST 106-2112-M-007 -006 -MY3.
The reported study was funded by RFBR and DFG 
according to the research project No 20-52-12009.
D. Bisht is financially supported by 
the Natural Science Foundation of China 
(NSFC-11590782, NSFC-11421303). 
This work has made use of data from the European Space Agency (ESA) mission
Gaia (http://www.cosmos.esa.int/gaia), processed by the Gaia Data
Processing and Analysis Consortium (DPAC, http://www.cosmos.
esa.int/web/gaia/dpac/ consortium). Funding for the DPAC has been
provided by national institutions, in particular the institutions 
participating in the Gaia Multilateral Agreement.
In this work we used a library for the Galactic dynamic (galpy)
in python created by J. Bovy.
We are grateful for the useful advice of J.~Bovy from the
Department of Astronomy and Astrophysics of the University of Toronto,
in particular, about using the galpy package.
LAMOST is operated and managed by the National Astronomical Observatories, 
Chinese Academy of Sciences.

%%%%%%%%%%%%%%%%%%%%%%%%%%%%%%%%%%%%%%%%%%%%%%%%%%%%%%%%%%%%%%%%%%%%%%%%%%%
\clearpage
\begin{table*}
%{\footnotesize
\tiny
\tabcolsep=0.07cm
\caption{Data for n=25 stars for which apexes were calculated. 
For stars in our ID column marked with an asterisk (*) 
the $V_r$ data were obtained from LAMOST DR5 catalog by adding 5 km s$^{-1}$, 
and for the other stars radial velocity is taken from Gaia DR2.}
\vspace{0.5cm}
\centering
\begin{tabular}{rrrrrrrrrrrrrrrr}
\hline\hline
\noalign{\smallskip}
our ID & P$_{\mu}$ & RA & DEC & A & D & Distance & Low bound & Upper bound & $V_r$ & $\sigma_{V_r}$ & $\mu_\alpha cos\delta$ & $\sigma_{\mu_\alpha cos\delta}$ &  $\mu_\delta$ & $\sigma_{\mu_\delta}$ \\
 & \% & J2000 & J2000 & $^\circ$ & $^\circ$ & pc & pc & pc & km s$^{-1}$ & km s$^{-1}$ & $mas~yr^{-1}$ & $mas~yr^{-1}$ & $mas~yr^{-1}$ & $mas~yr^{-1}$ \\
\hline
\noalign{\smallskip}

3966& 94.61& 92.14& 23.94& 90.98& $-$26.51& 1795.93&1664.22&1949.68& 13.36& 1.53&$-$0.045&0.074&$-$1.901&0.068\\
4943& 90.81& 91.86& 24.45& 86.18& $-$40.13& 4060.15&3448.68&4901.46& 16.21& 0.46&$-$0.150&0.078&$-$1.784&0.064\\
1116& 93.94& 91.77& 24.07& 91.36& $-$35.82& 3939.93&3514.36&4474.86& 22.39& 3.85&$-$0.014&0.051&$-$2.068&0.042\\
 679& 97.22& 91.80& 24.11& 87.90& $-$40.27& 5338.48&4362.64&6732.92& 23.71& 0.54&$-$0.113&0.075&$-$1.960&0.062\\
 998& 97.54& 91.91& 24.05& 85.18& $-$36.76& 4929.57&4326.82&5707.66& 24.74& 3.32&$-$0.206&0.049&$-$1.910&0.041\\
1203& 97.11& 91.92& 24.14& 89.33& $-$41.25& 5956.24&5143.55&7027.85& 25.77& 1.06&$-$0.075&0.043&$-$1.996&0.037\\
 314& 96.77& 91.87& 24.07& 82.12& $-$38.04& 5508.11&4769.56&6480.65& 25.95& 3.82&$-$0.290&0.051&$-$1.911&0.043\\
1708& 95.26& 91.71& 24.10& 86.04& $-$33.28& 4227.47&3203.96&5845.99& 27.11& 1.71&$-$0.209&0.121&$-$2.124&0.102\\
 383& 93.29& 91.87& 24.07& 90.21& $-$36.08& 4732.69&3811.66&6088.30& 27.54& 0.36&$-$0.058&0.082&$-$2.140&0.068\\
1263& 92.38& 91.92& 24.15& 88.11& $-$27.80& 4392.33&3779.98&5213.78& 29.34& 0.96&$-$0.135&0.065&$-$1.804&0.055\\
 120& 97.46& 91.84& 24.10& 85.47& $-$27.91& 4141.82&3632.79&4802.22& 31.39& 2.52&$-$0.257&0.049&$-$2.059&0.041\\
 305& 96.52& 91.84& 24.07& 85.34& $-$19.55& 3530.39&3132.85&4035.98& 33.42& 3.46&$-$0.297&0.066&$-$1.911&0.061\\
1225*&96.93& 91.88& 24.17& 89.58& $-$17.81& 3129.68&2749.93&3622.94& 31.63& 5.12&$-$0.110&0.063&$-$1.920&0.053\\
1509*&91.43& 91.75& 24.02& 86.98& $-$39.56& 4634.04&3649.69&6119.12& 23.92& 4.15&$-$0.158&0.082&$-$2.202&0.070\\
  96*&94.30& 91.86& 24.08& 82.12& $-$30.67& 4238.62&3546.98&5213.17& 28.91& 3.94&$-$0.370&0.077&$-$2.064&0.068\\
2389*&96.62& 91.94& 23.91& 89.59& $-$26.78& 1847.43&1705.05&2015.02& 13.76&11.46&$-$0.091&0.062&$-$1.921&0.052\\
 607*&96.64& 91.89& 24.12& 87.15& $-$26.38& 3884.08&3357.91&4586.58& 31.95& 8.72&$-$0.203&0.066&$-$2.111&0.056\\
 679*&97.22& 91.80& 24.11& 87.53& $-$44.16& 5338.48&4362.64&6732.92& 19.68& 3.19&$-$0.113&0.075&$-$1.960&0.062\\
 787*&97.88& 91.82& 24.14& 85.75& $-$36.10& 4428.83&3780.77&5310.08& 24.17& 4.61&$-$0.200&0.061&$-$2.026&0.052\\
3029*&97.89& 92.02& 24.28& 83.48& $-$40.14& 4831.88&4292.76&5512.59& 21.70& 4.52&$-$0.254&0.045&$-$2.008&0.038\\
4943*&90.81& 91.86& 24.45& 86.76& $-$35.35& 4060.15&3448.68&4901.46& 19.89& 3.72&$-$0.150&0.078&$-$1.784&0.064\\
3688*&96.54& 92.00& 24.35& 87.27& $-$43.47& 4861.61&3747.84&6553.76& 18.92&11.63&$-$0.131&0.117&$-$2.024&0.099\\
3765*&96.99& 91.53& 24.07& 81.64& $-$49.67& 6033.00&4439.32&8329.23& 16.04& 2.77&$-$0.230&0.114&$-$1.978&0.092\\
5394*&99.80& 92.16& 24.35& 89.91&    16.97& 2388.29&1578.92&4004.46& 39.03& 5.33&$-$0.131&0.359&$-$0.446&0.304\\
1708*&95.26& 91.71& 24.10& 86.63& $-$28.42& 4227.47&3203.96&5845.99& 32.53& 3.52&$-$0.209&0.121&$-$2.124&0.102\\
\hline
\label{tabapex}
\end{tabular}
\end{table*}
%%%%%%%%%%%%%%%%%%%%%%%%%%%%%%%%%%%%%%%%%%%%%%%%%%%%%%%%%%%%%%%%%%%%%%%%%%%

%%%%%%%%%%%%%%%%%%%%%%%%%%%%%%%%%%%%%%%%%%%%%%%%%%%%%%%%%%%%%%%%%%%%%%%%%%%
\clearpage
\begin{table*}
\footnotesize
\tabcolsep=0.1cm
\caption{Parameters of NGC~2158 orbit integration.}
%Please note that we only use the mean radial velocity of 12 Gaia-DR2 stars
%because the LAMOST DR5 stars had higher errors that could affect our results.
\vspace{0.5cm}
\centering
\begin{tabular}{ccccccccc}
\hline\hline
\noalign{\smallskip}
\textbf{Input parameters} \\
$\mu_{\alpha}cos\delta$&$\mu_{\delta}$&d&RA J2000&DEC J2000&$V_{r}$\\
mas yr$^{-1}$ & mas yr$^{-1}$ & kpc & [ $^\circ$ ] & [ $^\circ$ ] & km s$^{-1}$\\
\hline
\noalign{\smallskip}
$-$0.203 & $-$1.992 & 4.69 & 91.862 & 24.099 & 25.1\\
\\
\hline
\hline
\noalign{\smallskip}
\textbf{Output parameters} \\
Apocenter & Pericenter & $e$ & $Z_{max}$ & $U$ & $V$ & $W$ & $T_{r}$ \\
kpc & kpc & & pc & km s$^{-1}$ & km s$^{-1}$ & km s$^{-1}$ & Gyr \\
\hline
\noalign{\smallskip}
12.8$\pm$0.03 & 11.1$\pm$0.05 & 0.07$\pm$0.003 & 529$\pm$7 & $-$21.5$\pm$1.7 & $-$39.2$\pm$1.7 & $-$24.6$\pm$1.7 & $\sim$0.2 \\

\hline
\label{taborbitinput}
\end{tabular}
\end{table*}

%%%%%%%%%%%%%%%%%%%%%%%%%%%%%%%%%%%%%%%%%%%%%%%%%%%%%%%%%%%%%%%%%%%%%%%%%%%
%%%%%%%%%%%%%%%%%%%%%%%%%%%%%%%%%%%%%%%%%%%%%%%%%%%%%%%%%%%%%%%%%%%%%%%%%%%
%%%%%%%%%%%%%%%%%%%%%%%%%%%%%%%%%%%%%%%%%%%%%%%%%%%%%%%%%%%%%%%%%%%%%%%%%%%
%%%%%%%%%%%%%%%%%%%%%%%%%%%%%%%%%%%%%%%%%%%%%%%%%%%%%%%%%%%%%%%%%%%%%%%%%%%
%%%%%%%%%%%%%%%%%%%%%%%%%%%%%%%%%%%%%%%%%%%%%%%%%%%%%%%%%%%%%%%%%%%%%%%%%%%
%%%%%%%%%%%%%%%%%%%%%%%%%%%%%%%%%%%%%%%%%%%%%%%%%%%%%%%%%%%%%%%%%%%%%%%%%%%
%%%%%%%%%%%%%%%%%%%%%%%%%%%%%%%%%%%%%%%%%%%%%%%%%%%%%%%%%%%%%%%%%%%%%%%%%%%
%\begin{figure}
%%\centering
%\includegraphics[width=\textwidth]{fig_vpdbinned_n2158.ps}
%\caption{
%We show here the CMDs and VPDs according to different magnitude bins.
%The CMD in left panel is for all the stars plotted in the VPDs
%and in the right panels, CMDs for only the stars within 
%the circles of the VPDs
%From brighter to fainter bins, the PM errors increase from  
%0.5 mas~yr$^{-1}$ to 1.0 mas~yr$^{-1}$
%and the radii of the VPD circles increase from 
%0.3 mas~yr$^{-1}$ to 0.6 mas~yr$^{-1}$.
%}
%\label{figvpd_binned}
%\end{figure}
%%%%%%%%%%%%%%%%%%%%%%%%%%%%%%%%%%%%%%%%%%%%%%%%%%%%%%%%%%%%%%%%%%%%%%%%%%%
%%%%%%%%%%%%%%%%%%%%%%%%%%%%%%%%%%%%%%%%%%%%%%%%%%%%%%%%%%%%%%%%%%%%%%%%%%
%\begin{figure}
%\includegraphics[width=\textwidth]{figmp_n2158.ps}
%\caption{
%The cluster membership probabilities for stars in 
%the region of NGC 2158 with Gaia $G$ magnitude.
%}
%\label{figmp}
%\end{figure}
%%%%%%%%%%%%%%%%%%%%%%%%%%%%%%%%%%%%%%%%%%%%%%%%%%%%%%%%%%%%%%%%%%%%%%%%%%%
%%%%%%%%%%%%%%%%%%%%%%%%%%%%%%%%%%%%%%%%%%%%%%%%%%%%%%%%%%%%%%%%%%%%%%%%%%%
%%%%%%%%%%%%%%%%%%%%%%%%%%%%%%%%%%%%%%%%%%%%%%%%%%%%%%%%%%%%%%%%%%%%%%%%%%%
%%%%%%%%%%%%%%%%%%%%%%%%%%%%%%%%%%%%%%%%%%%%%%%%%%%%%%%%%%%%%%%%%%%%%%%%%%%
%%%%%%%%%%%%%%%%%%%%%%%%%%%%%%%%%%%%%%%%%%%%%%%%%%%%%%%%%%%%%%%%%%%%%%%%%%%
%%%%%%%%%%%%%%%%%%%%%%%%%%%%%%%%%%%%%%%%%%%%%%%%%%%%%%%%%%%%%%%%%%%%%%%%%%%
%%%%%%%%%%%%%%%%%%%%%%%%%%%%%%%%%%%%%%%%%%%%%%%%%%%%%%%%%%%%%%%%%%%%%%%%%%%
%%%%%%%%%%%%%%%%%%%%%%%%%%%%%%%%%%%%%%%%%%%%%%%%%%%%%%%%%%%%%%%%%%%%%%%%%%%
%%%%%%%%%%%%%%%%%%%%%%%%%%%%%%%%%%%%%%%%%%%%%%%%%%%%%%%%%%%%%%%%%%%%%%%%%%%
%%%%%%%%%%%%%%%%%%%%%%%%%%%%%%%%%%%%%%%%%%%%%%%%%%%%%%%%%%%%%%%%%%%%%%%%%%%
%%%%%%%%%%%%%%%%%%%%%%%%%%%%%%%%%%%%%%%%%%%%%%%%%%%%%%%%%%%%%%%%%%%%%%%%%%%
%%%%%%%%%%%%%%%%%%%%%%%%%%%%%%%%%%%%%%%%%%%%%%%%%%%%%%%%%%%%%%%%%%%%%%%%%%%
%%%%%%%%%%%%%%%%%%%%%%%%%%%%%%%%%%%%%%%%%%%%%%%%%%%%%%%%%%%%%%%%%%%%%%%%%%%
%%%%%%%%%%%%%%%%%%%%%%%%%%%%%%%%%%%%%%%%%%%%%%%%%%%%%%%%%%%%%%%%%%%%%%%%%%%
%%%%%%%%%%%%%%%%%%%%%%%%%%%%%%%%%%%%%%%%%%%%%%%%%%%%%%%%%%%%%%%%%%%%%%%%%%%

%%%%%%%%%%%%%%%%%%%%%%%%%%%%%%%%%%%%%%%%%%%%%%%%%%%%%%%%%%%%%%%%%%%%%%%%%%%
\clearpage
%%%%%%%%%%%%%%%%%%%%%%%%%%%%%%%%%%%%%%%%%%%%%%%%%%%%%%%%%%%%%%%%%%%%%%%%%%%

%%%%%%%%%%%%%%%%%%%%%%%%%%%%%%%%%%%%%%%%%%%%%%%%%%%%%%%%%%%%%%%%%%%%%%%%%%%
\clearpage
%%%%%%%%%%%%%%%%%%%%%%%%%%%%%%%%%%%%%%%%%%%%%%%%%%%%%%%%%%%%%%%%%%%%%%%%%%%

%%%%%%%%%%%%%%%%%%%%%%%%%%%%%%%%%%%%%%%%%%%%%%%%%%%%%%%%%%%%%%%%%%%%%%%%%%%
\clearpage
%%%%%%%%%%%%%%%%%%%%%%%%%%%%%%%%%%%%%%%%%%%%%%%%%%%%%%%%%%%%%%%%%%%%%%%%%%%

%%%%%%%%%%%%%%%%%%%%%%%%%%%%%%%%%%%%%%%%%%%%%%%%%%%%%%%%%%%%%%%%%%%%%%%%%%%
\clearpage
%%%%%%%%%%%%%%%%%%%%%%%%%%%%%%%%%%%%%%%%%%%%%%%%%%%%%%%%%%%%%%%%%%%%%%%%%%%

%%%%%%%%%%%%%%%%%%%%%%%%%%%%%%%%%%%%%%%%%%%%%%%%%%%%%%%%%%%%%%%%%%%%%%%%%%%
\clearpage
%%%%%%%%%%%%%%%%%%%%%%%%%%%%%%%%%%%%%%%%%%%%%%%%%%%%%%%%%%%%%%%%%%%%%%%%%%%

\end{document}